\documentclass[pdflatex,sn-mathphys-num]{sn-jnl}

\usepackage{graphicx}%
\usepackage{multirow}%
\usepackage{amsmath,amssymb,amsfonts}%
\usepackage{amsthm}%
\usepackage{mathrsfs}%
\usepackage[title]{appendix}%
\usepackage{xcolor}%
\usepackage{textcomp}%
\usepackage{manyfoot}%
\usepackage{booktabs}%
\usepackage{algorithm}%
\usepackage{algorithmicx}%
\usepackage{algpseudocode}%
\usepackage{listings}%

\unnumbered

\begin{document}

\title[Article Title]{GHz-speed wavefront shaping metasurface modulators enabled by resonant electro-optic nanoantennas}

\author*[1]{\fnm{Sahil} \sur{Dagli}}\email{daglis@stanford.edu}
\author[1]{\fnm{Jiyong} \sur{Shim}}
\author[1]{\fnm{Hamish} \sur{Carr Delgado}}
\author[1]{\fnm{Halleh B.} \sur{Balch}}
\author[1]{\fnm{Sajjad} \sur{Abdollahramezani}}
\author[1]{\fnm{Chih-Yi} \sur{Chen}}
\author[1]{\fnm{Varun} \sur{Dolia}}
\author[1]{\fnm{Elissa} \sur{Klopfer}}
\author[2]{\fnm{Jefferson} \sur{Dixon}}
\author[1]{\fnm{Jack} \sur{Hu}}
\author[1]{\fnm{Babatunde} \sur{Ogunlade}}
\author[3]{\fnm{Jung-Hwan} \sur{Song}}
\author[4]{\fnm{Mark L.} \sur{Brongersma}}
\author*[5]{\fnm{David} \sur{Barton}}\email{dbarton@northwestern.edu}
\author*[1]{\fnm{Jennifer A.} \sur{Dionne}}\email{jdionne@stanford.edu}

\affil*[1]{\orgdiv{Department of Materials Science and Engineering}, \orgname{Stanford University}, \orgaddress{\city{Stanford}, \state{California}, \country{USA}}}

\affil[2]{\orgdiv{Department of Mechanical Engineering}, \orgname{Stanford University}, \orgaddress{\city{Stanford}, \state{California}, \country{USA}}}

\affil[3]{\orgdiv{Department of Electrical and Computer Engineering}, \orgname{National University of Singapore}, \orgaddress{\city{Singapore}, \country{Singapore}}}

\affil[4]{\orgdiv{Geballe Laboratory for Advanced Materials}, \orgname{Stanford University}, \orgaddress{\city{Stanford}, \state{California}, \country{USA}}}

\affil[5]{\orgdiv{Department of Materials Science and Engineering}, \orgname{Northwestern University}, \orgaddress{\city{Evanston}, \state{Illinois}, \country{USA}}}

\abstract{Electrically tunable metasurfaces that control the amplitude and phase of light through biasing of nanoscale antennas present a route to compact, sub-micron thick modulator devices. However, most platforms face limitations in bandwidth, absolute optical efficiency, and tuning response. Here, we present electro-optically tunable metasurfaces capable of both GHz amplitude modulation and transmissive wavefront shaping in the telecom range. Our resonant electro-optic nanoantenna design consists of a silicon nanobar atop thin-film lithium niobate, with gold electrodes. The silicon nanobar is a periodically perturbed optical waveguide that supports high quality factor (Q $>$ 1000) guided mode resonances excited with free space light. Applying a voltage bias to the lithium niobate tunes its refractive index, modulating the resonant behavior of the silicon nanobar through evanescent mode overlap. We demonstrate an absolute transmittance modulation of 7.1\% with $\pm$5 V applied voltage, and show the dependence of this modulation behavior on the resonance quality factor. We additionally study the electrode limitations on modulation bandwidth, demonstrating bandwidths exceeding 800 MHz. Finally, we show how this resonant antenna platform can be used to design wavefront shaping metasurfaces. We demonstrate a beamsplitting metasurface device, whose diffraction efficiency can be modulated with a bandwidth of 1.03 GHz. The high-speed modulation and wavefront control capabilities of this platform provide a foundation for compact, high bandwidth free space communications and sensing devices.}


\maketitle

\section{Introduction}\label{sec1}

High speed electro-optic modulation of photonic devices is a critical functionality for high bandwidth communications platforms\cite{Liu2015}. These platforms rely on the ability to encode information in the amplitude, phase, or polarization of light, and to reconfigure these attributes at will. While considerable attention has been given to modulators for photonic integrated circuits\cite{Sinatkas2021}, modulation of free-space light is of increasing importance. Compact and reconfigurable free-space optical systems could advance applications spanning light detection and ranging (LiDAR)\cite{Kim2021}, light-fidelity (LiFi) for short-range, fast, and secure communications\cite{Abumarshoud2022,Thureja2022}, augmented/virtual reality (AR/VR)\cite{Gopakumar2024,Song2021}, holography\cite{Huang2018,Xiong2023}, and entangled quantum networks\cite{Solntsev2021}. Optical metasurfaces enable compact, multifunctional free-space optical systems using arrays of specially designed nanoantennas that can perform the appropriate transfer function with subwavelength control\cite{Kuznetsov2024}. Dynamic, electro-optic control of the constituent nanoantennas in a metasurface\cite{Jung2024,Shaltout2019f,Ha2024} has been demonstrated using materials such as liquid crystals\cite{Li2019s,Moitra2023,Cai2024}, phase change materials\cite{Zhang2021,Wang2020a,Abdollahramezani2022,Fang2024}, and doped semiconductors\cite{Huang2016d,Shirmanesh2020d,Park2020e,Kim2017} to tune the device’s function. However, these approaches face tradeoffs in bandwidth, absorption, and index contrast, requiring new materials and design strategies to achieve full wavefront tunability at speeds exceeding 1 GHz.

Electro-optic materials provide a platform to enable high speed, low loss modulation. Lithium niobate (LNO) is a well-known electro-optic material that exhibits a strong Pockels effect ($r_{33}$ = 31 pm/V) due to its noncentrosymmetric crystal structure\cite{Weis1985c,Boes2023}. Since the Pockels effect is a nonlinear effect that modifies a material’s susceptibility without large structural changes, the material can be switched at very high speeds ($>$100 GHz in integrated photonic devices)\cite{Wang2018i}. It additionally features a wide band gap of $\sim$4 eV, making it transparent in the visible and near-infrared and therefore applicable for a variety of devices. LNO offers robustness in extreme conditions as well as reliable film quality with uniform ferroelectric domain orientation from scaled production compared to alternative electro-optic materials\cite{Gu2023}, such as electro-optic polymers\cite{Benea-Chelmus2022,Zheng2024,Lin2025} and barium titanate\cite{Weigand2024,Abel2019}. Finally, the commercial availability of thin film LNO on insulator (LNOI) has led to the widespread development of LNO-based integrated photonic and metasurface devices with electro-optic tunability\cite{Li2020a,Zhang2021a,Weigand2021,Weiss2022,Damgaard-Carstensen2023,Zhang2019a,Stokowski2024,Fedotova2022,Chen2025}. While low voltage and high bandwidth modulators can be fabricated, they are typically designed with long optical path lengths (V$\pi$L $\sim$ 1 V-cm)\cite{Wang2018i,Kharel2021,Li2023b,Mercante2016} to maximize the electro-optic response from the modest index change in LNO. Integrating LNO into metasurfaces has therefore been challenged by the prohibitively large voltages required to generate the needed phase shifts required for reconfigurability\cite{Weigand2021}. 

Engineering the optical modes of nanoantennas to exhibit high-quality factor (high-Q) resonances can potentially enable compact, free-space reconfigurable photonic systems for wavefront shaping\cite{Barton2020,Hail2023,So2023}. By applying an electric field across the nanoantenna, the resonant wavelength shifts, leading to significant variation in the scattered phase or amplitude from the narrow bandwidth resonance. This principle has been applied in high-Q metasurfaces supporting bound-in-continuum (BIC) modes, though these structures have nonlocal mode volumes distributed across both in-plane dimensions\cite{Benea-Chelmus2022,DiFrancescantonio2024,Damgaard-Carstensen2024,Overvig2020,Shastri2023,Koshelev2018b}. Given that wavefront shaping typically requires subwavelength control of phase and/or amplitude in at least one dimension, these high-Q structures have been limited in their functionality.

Here, we demonstrate metasurfaces that enable both efficient wavefront modulation through resonant optical scattering and individual nanoantenna addressability, providing a path to spatial light modulators with a nanoscale pixel pitch. Our device uses a hybrid materials platform of silicon on lithium niobate, taking advantage of the high index and lossless properties of silicon in the telecom wavelength range and the electro-optic Pockels effect in lithium niobate. We design subwavelength (400 nm width) nanobars of 300 nm thick silicon atop a 300 nm lithium niobate thin film. By introducing subtle geometric perturbations into each constituent nanobar, we generate leaky guided mode resonances that couple to normally-incident free-space light. These guided mode resonances exhibit mode quality factors exceeding 1000 in both theoretical and experimental demonstrations, and we demonstrate that the amplitude and phase of scattered light from these resonances can be modulated with small voltage inputs. We first present a metasurface device that acts as an amplitude modulator, with peak transmission modulation of 7.1\% (absolute) with $\pm$5 V applied to the metasurface in experiments. This resonant shifting efficiency of 0.71\%/V is a significant improvement for devices with this level of absolute tunability and bandwidth. Importantly, we find that the current modulation bandwidth of our devices is not limited by the quality factor of the metasurface; by miniaturizing the metasurface further, we enable modulation bandwidths beyond 800 MHz by reducing the overall capacitance of the electrodes. Building on this, we finally show the general applicability of this high-Q Si on LNO nanoantenna platform for wavefront shaping by developing a reconfigurable high-Q beamsplitter. This design features a 3 dB bandwidth greater than 1 GHz. Our resonant metasurface platform provides a general foundation for combining high frequency modulation with phase gradient enabled wavefront shaping, whereby each nanoantenna can be efficiently modulated with an electric field. This can enable a new suite of efficient and high speed reconfigurable optical elements with potential applications in ultrathin dynamic optics, free-space communications, modulation and sensing. 

\section{High-Q, electro-optic nanoantenna platform}\label{sec2}

Figure 1a shows our metasurface design. Our device uses 300 nm tall silicon nanobars atop a 300 nm LNOI substrate, with gold electrodes alongside the nanobars directly in contact with the LNO layer. Light transmitted through the metasurface is modulated in intensity through coupling with the resonant electro-optic nanoantennas. Figure 1b shows a schematic of our electro-optic nanoantenna, highlighting key dimensions of the unit cell. The Si nanobar is a periodically perturbed optical waveguide structure, elongated in one dimension and 400 nm in width. Each Si nanobar supports bound waveguide modes that have higher momentum than free space light (see Supplementary note 1). Periodic perturbations (described by their depth, $d$, width, $w$, of 100 nm, and period, $p$, of 680 nm) along each nanobar allow normally incident free space light to couple to waveguide modes, resulting in guided mode resonances (GMRs) with high quality factors in the far-field response. These nanoantennas scatter light with a dipole-like profile, with an amplitude and phase that can be controlled via the applied electric field due to the resonant mode overlap with the LNO layer underneath. 

A high-Q resonance provides a sharp spectral feature over a narrow bandwidth, enabling modulation over a wide dynamic range with a minimal applied voltage. Our modulation scheme is shown in figure 1c. Application of an electric bias modulates the refractive index in the LNO layer, shifting the resonant wavelength of the guided mode. For a single illumination wavelength, the spectral shift of the guided mode resonance results in a modulation of the transmitted intensity. The efficiency of this process is related to the derivative of the asymmetric Lorentzian resonance (Fig. 1c, bottom panel), with maximum modulation efficiency near the full-width half-max of the resonance where the slope of the asymmetric Lorentzian lineshape is the largest. The simulated results in figure 1c are shown for a nanoantenna design with a perturbation depth $d$ = 40 nm, a quality factor of 4150, and an applied voltage of $\pm$5 V. 

High-Q metasurface designs that support subwavelength mode confinement in one dimension while allowing extended, guided mode propagation in the other dimension allow for multiplexing nanoantenna function within a single metasurface, a key advantage over designs where the mode is distributed across both dimensions of the metasurface\cite{Benea-Chelmus2022,Damgaard-Carstensen2023}. These high-Q structures can be patterned and etched into silicon thin films, taking advantage of robust fabrication procedures as well as the high refractive index properties silicon has to offer\cite{Witmer2017d,Weigel2016d,Chiles2014b}. With these design considerations in mind, we fabricated a variety of devices using single crystal thin film direct bonding, electron beam lithography, reactive ion etching, and electrode metallization (see methods). Figure 1d shows a microscopy image of several test devices with metasurface dimensions 160 $\times$ 160 $\mu$m, each with 100 $\times$ 50 $\mu$m probe pads for AC voltage application, highlighting the potential for multiplexing capabilities of these metasurface devices. Figure 1e shows a false-color scanning electron microscope image of a fully fabricated device, where each nanoantenna can be individually addressed.

\begin{figure}[h]
\centering
\includegraphics[width=\textwidth]{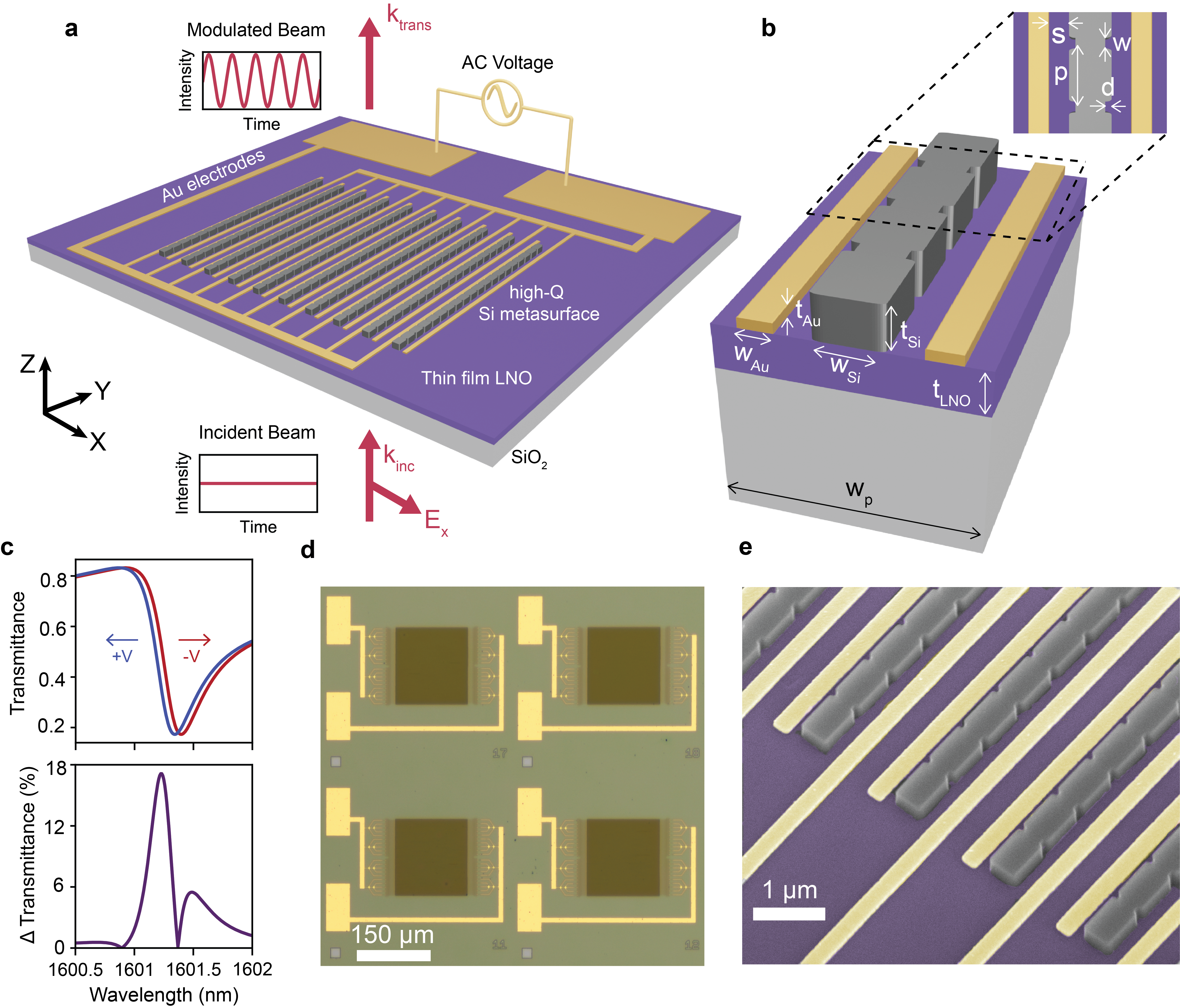}
\caption{(a) Schematic of metasurface device. Incident light ($k_{inc}$) with an x-polarized electric field transmits ($k_{trans}$) through the metasurface, upon which the intensity of the transmitted light is modulated. (b) Schematic of nanoantenna unit cell, consisting of silicon nanoantenna, LNO thin film, and gold electrodes, where $p$ = 680 nm, $w$ = 100 nm, $d$ = 40-100 nm, $s$ = 50-350 nm, $w_{Si}$ = 400 nm, $t_{Si}$ = 300 nm, $t_{LNO}$ = 300 nm, $t_{Au}$ = 70 nm, $w_{Au}$ = 200 nm, and $w_p$ = 1600 nm. (c) Principle of operation: the shift in resonant wavelength due to the change in refractive index of the LNO results in the intensity of the transmitted light being modulated. Simulated results are from a metasurface with a perturbation depth, d, of 40 nm, and electrode spacing, s, of 300 nm, applying $\pm$5 V. (d) Optical micrograph of the fabricated metasurface devices. (e) False-color scanning electron micrograph of a fabricated device.}\label{fig1}
\end{figure}

\section{Design and characterization of high-Q nanoantennas}\label{sec3}

The generated GMR is primarily localized to the silicon nanobar with evanescent overlap into the LNO layer (9.6\% of the mode), as shown in figure 2a for a $d$ = 40 nm nanoantenna. The LNO is x-cut, such that a horizontally (x-direction) polarized electric dipole both exhibits point-like, dipolar radiation for wavefront shaping while also maximizing electro-optic overlap with the strong Pockels coefficient ($r_{33}$ = 31 pm/V, compared to $r_{13}$ = 9 pm/V). Critically, electrodes around each individual nanoantenna enable, in principle, nearly arbitrary wavefront shaping capabilities in one dimension\cite{Klopfer2022}. Applying a voltage across these electrodes induces a permittivity change in the LNO through the electro-optic effect:

\begin{equation}
    \Delta\epsilon_{xx} = -r_{33}n_e^4E_x^{app}
\end{equation}
\begin{equation}
    \Delta\epsilon_{yy} = \Delta\epsilon_{zz} = -r_{13}n_o^4E_x^{app}
\end{equation}

Here, $n_e$ = 2.14 and is the extraordinary refractive index of LNO,  $n_o$ = 2.21 and is the ordinary refractive index of LNO, and $E_x^{app}$ is the applied electric field from the voltage bias across the electrodes. Our simulated results show a resonant wavelength shift of 5.4-5.9 pm/V, depending on the perturbation depth (see supplementary note 2). Figure 2b shows the simulated applied electric field profile from the voltage bias. With the relatively high dielectric constant of LNO ($\sim$30)\cite{Chelladurai2024}, directly contacting the material is critical for sufficient electric field to enable effective overlap with the optical mode at low voltage frequencies\cite{Witmer2016d}. This additionally maximizes electric field strengths without substantially modifying the quality factor from optical absorption. Together, this composite structure provides a nominally optimized geometry compared to a purely LNO device layer\cite{Klopfer2022,Barton2021d} (see supplementary note 7).

Varying the perturbation depth, $d$, controls the coupling of free space light to the guided mode resonance, and thus the quality factor\cite{Lawrence2018i,Overvig2018,Kim2019k}. Figure 2c shows simulated transmittance across the resonance for varying perturbation depths. A higher Q resonance corresponds to a sharper amplitude contrast across the resonance, which will provide a larger amplitude and phase shift response with applied voltage. Indeed, while other designs prioritize mode confinement in the LNO layer that may experience high absorption from the optical mode being close to the electrodes, we achieve high optical efficiency with the high index Si confining the mode sufficiently far from the electrodes to minimize the absorptance (see supplementary note 3). We fabricate a series of metasurfaces with the same perturbation depths as in figure 2c and characterize them using a home-built transmission microscope setup, with experimental results shown in figure 2d. These devices show the same trend of increasing quality factor with decreasing perturbation depth, with quality factors ranging from 500 to 2200. The smaller amplitude contrast seen in experimental results is attributed to other dominating loss mechanisms, such as multiple scattering channels opened by fabrication processes causing nonuniformities. In supplementary note 4, we fabricate and characterize a series of metasurfaces varying the perturbation period, $p$, as well. We additionally measure devices without electrodes to confirm that the metal does not strongly impact the resonant response of our devices. Figure 2e shows the simulated and experimental quality factors of metasurfaces with and without electrodes. In both simulations and experiments, the introduction of electrodes at the appropriate position does not significantly diminish the quality factor of our nanoantennas. 

The amplitude modulation performance is a key figure of merit, determined by the combination of a resonant nanoantenna’s quality factor and resonant shifting with a given voltage. We quantify this figure of merit, defined as the maximum percent change in transmittance per volt, with simulated results in figure 2f for a device with a perturbation depth $d$ = 40 nm. The placement of the electrodes and their separation from the resonant Si nanoantenna affects the modulator performance. Moving the electrodes further away from the antenna moderately increases the Q as absorption losses decrease. However, the local electric field from the applied voltage decreases with electrode spacing, creating a tradeoff with tuning efficiency, or change in resonant wavelength per volt (see figure S8). For this given geometry, we find that there is an optimal electrode spacing 200-300 nm away from the Si nanobar. This distance is within standard fabrication tolerances, providing a promising design for consistent operation. We fabricate devices in the following section with a design of $s$ = 300 nm. Additionally, it highlights the tradeoffs in optimizing for modulator performance; in these resonant devices, the tuning efficiency is less important than transmission contrast for maximizing modulation amplitude. Having higher Q resonances, which maximize the narrowband amplitude contrast, can show better modulation performance even with a modest tuning efficiency. 

\begin{figure}[h]
\centering
\includegraphics[width=\textwidth]{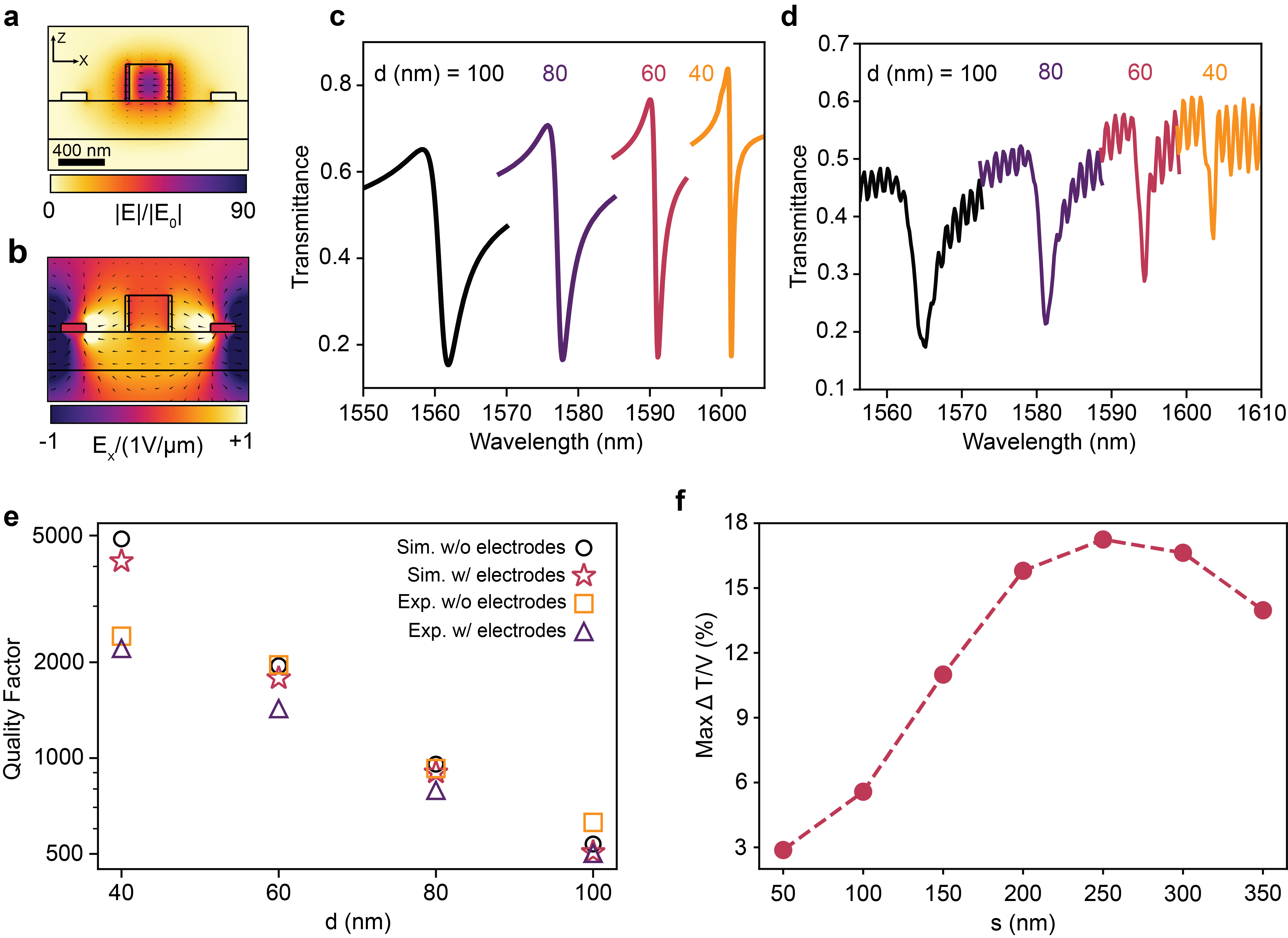}
\caption{(a) Enhanced nearfield of resonances in the Si nanoantenna for a design with $d$ = 40 nm at 1601.3 nm. (b) Electric field profile of applied voltage from the gold electrodes. (c) Simulated and (d) experimental spectra from high-Q metasurfaces with electrodes, varying perturbation depth, $d$. (e) Simulated and experimental quality factors from metasurfaces with and without electrodes. (f) Simulated maximum change in transmittance per volt from a $d$ = 40 nm metasurface as a function of electrode spacing, $s$.}\label{fig2}
\end{figure}

\section{Resonance-dependent electro-optic modulation}\label{sec4}

With our optimized device geometry fabricated, we demonstrate electro-optic amplitude modulation by applying an AC voltage to the metasurface. Figure 3a shows the 1 MHz, $\pm$5 V voltage as a function of time (top), with the photoresponse (bottom) from our detector showing the resulting modulated intensity of the light after transmitting through the metasurface. The laser intensity is modulated at the same frequency as the applied voltage. The phase delay results from a positive voltage decreasing the transmittance at this wavelength (see Supplementary note 8). As discussed earlier, decreasing $d$ increases the Q of the resonance, leading to a higher modulation signal for a given applied voltage. Figure 3b shows simulated results for modulation amplitude, with a maximum change in transmittance of 1.9, 3.3, 6.7, and 16.6\% for metasurfaces with $d$ = 100, 80, 60, and 40 nm respectively.

We fabricate metasurfaces that are 160 $\times$ 160 $\mu$m in area to verify these results. While applying a 1 MHz, $\pm$5 V voltage to the device, we measure the modulated laser intensity and normalize that signal to the background transmittance to determine the modulation signal as a percent change in transmittance. Figure 3c shows the experimental data, where we observe maximum transmittance modulation of 1.6, 2.6, 3.9, and 7.1\% for metasurfaces with $d$ = 100, 80, 60, and 40 nm respectively. We attribute the difference between simulated and experimental results to a smaller resonance amplitude contrast for the experimentally fabricated devices. We see additional peaks in the modulation signal throughout the spectrum, which are a result of the Fabry-Perot response of the LNO substrate interfering with the modulation signal. Focusing on the results from the metasurface with $d$ = 40 nm, figure 3d shows the modulation efficiency while varying the applied RF voltage from V = $\pm$1-5 V. We see a stable and linear response, as shown in figure 3e, indicating that the resonance is not susceptible to thermal shifts or other sources of drift in this voltage range and frequency. The slope of the resonant lineshape is roughly linear around the full-width half-max, and we expect to see a diminishing, nonlinear increase in the modulation response as the voltage is further increased and more of the resonance is swept out during modulation.

\begin{figure}[h]
\centering
\includegraphics[width=\textwidth]{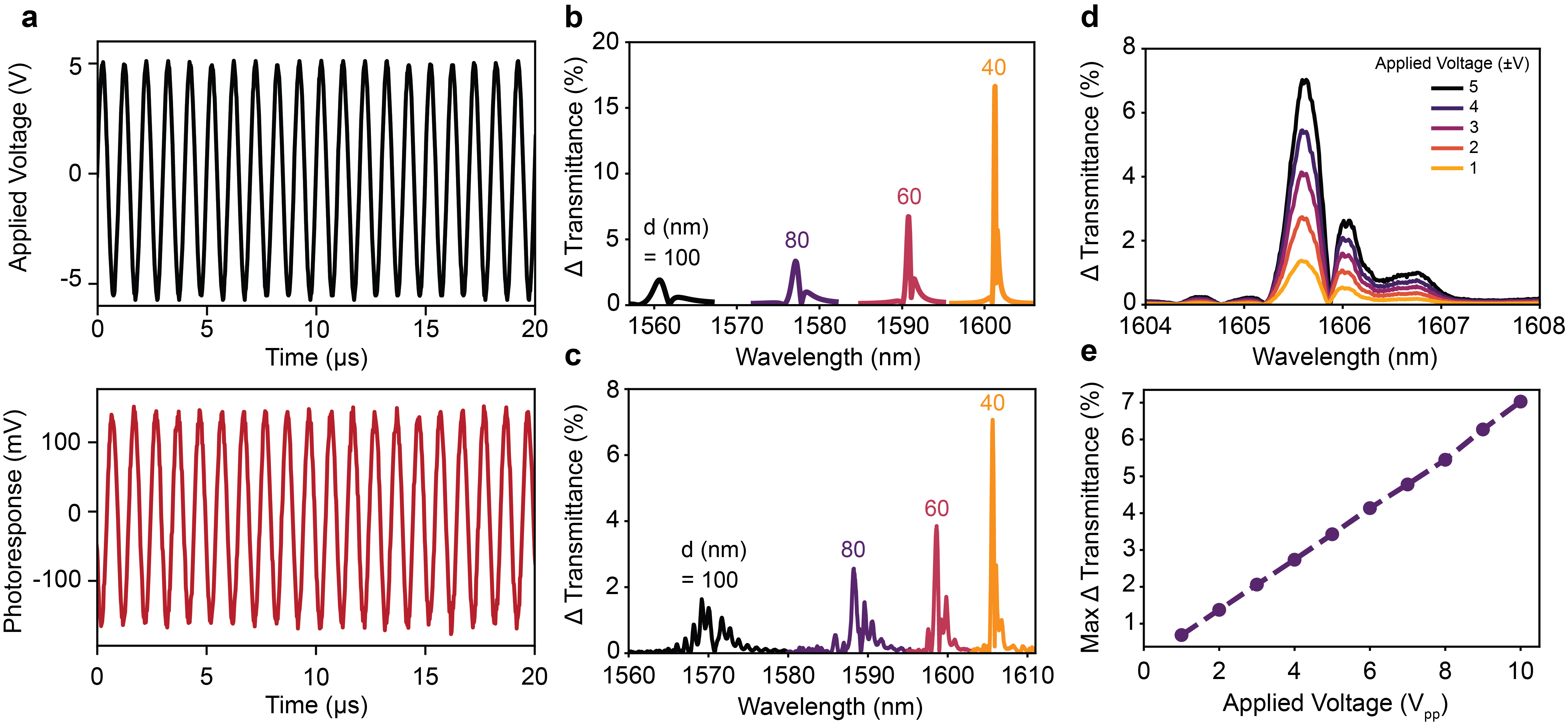}
\caption{(a) (top) Applied voltage and (bottom) modulated laser intensity (photoresponse) as a function of time for the $d$ = 40 nm metasurface device at 1605.63 nm. (b) Simulated and (c) experimental modulation amplitude from metasurfaces with varying perturbation depth, $d$. (d) Modulation amplitude as a function of wavelength with varying applied voltage for the $d$ = 40 nm metasurface device. (e) Maximum modulation amplitude as a function of applied voltage for the $d$ = 40 nm metasurface device.}\label{fig3}
\end{figure}

\section{High bandwidth modulation enabled by miniaturized metasurfaces}\label{sec5}

We determine the modulation bandwidth of these metasurfaces by sweeping the frequency of the applied voltage from 500 kHz - 1.6 GHz. We apply a $\pm$1 V voltage to our devices for these measurements, to minimize thermal drift at high frequencies. The modulation response is normalized to the value recorded at 500 kHz. Recognizing that the metasurface electrodes have a capacitance and resistance that will limit the high frequency modulation response, we fabricate a series of metasurfaces with varying area to investigate the limiting factors in our metasurface design. Decreasing the area of the device will decrease the capacitance of the electrodes, thus increasing the RC-limited modulation bandwidth of the device. Each metasurface is a square array, and the width is varied from 160 - 40 $\mu$m. Figure 4a shows the modulation as a function of frequency for each metasurface, with the inset showing optical microscope images of fabricated metasurfaces of 160 (left) and 40 (right) $\mu$m width. 

The metasurfaces all exhibit a stable modulation response from 500 kHz - 100 MHz, after which the modulation amplitude starts to decrease. We define the 3 dB bandwidth as the frequency where the modulation response first drops below -3 dB. Figure 4b shows the extracted bandwidth values for each metasurface. The measured bandwidths are 340, 470, 700, and 890 MHz for metasurfaces with widths of 160, 120, 80, and 40 $\mu$m respectively. We compare these results to expected bandwidth values based on capacitance calculations in supplementary note 5. This modulation bandwidth is not limited by the Q of the resonant metasurface, as the Q decreases with smaller width (see figure S9). Further design improvements, such as combining the GMR structure with a photonic crystal cavity to taper the GMR boundary and prevent leakage, can improve the Q as desired\cite{Dolia2024}.

\begin{figure}[h]
\centering
\includegraphics[width=\textwidth]{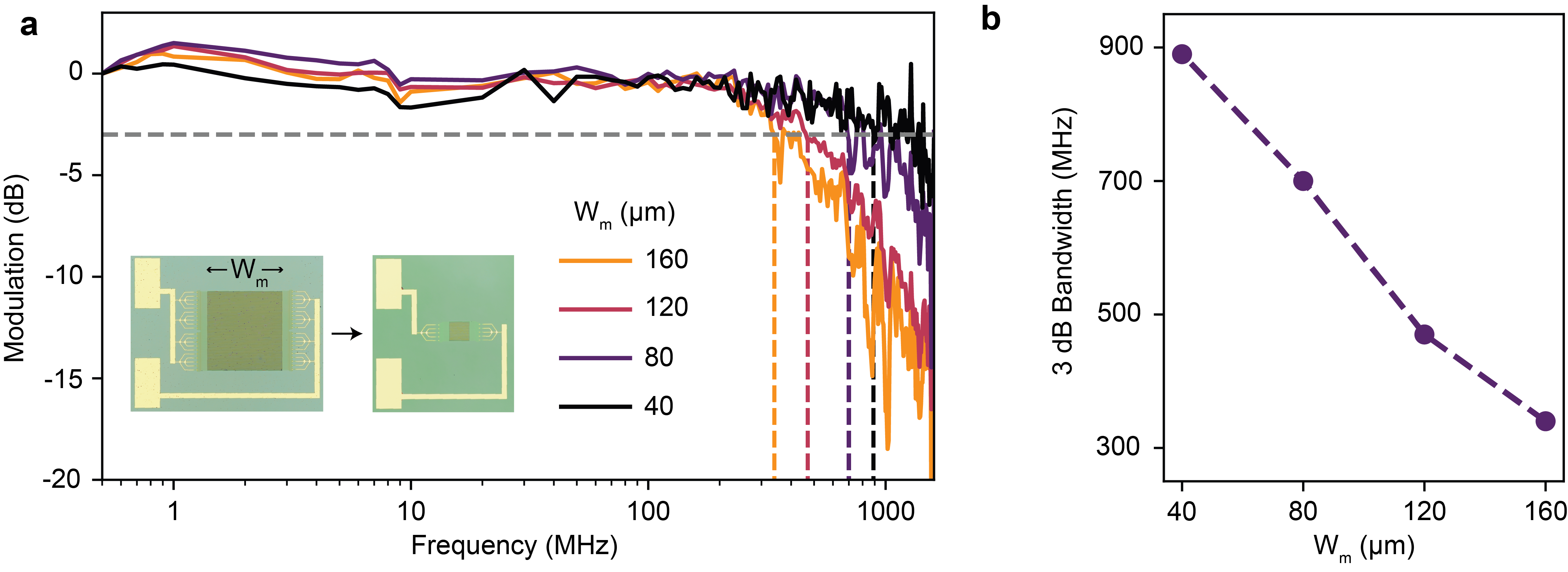}
\caption{(a) Modulation as a function of frequency, normalized to the magnitude at 500 kHz, for devices varying the metasurface width from 40-160 $\mu$m. Inset shows optical microscope images of devices with metasurface width of 160 (left) and 40 (right) $\mu$m. The grey dashed line indicates the 3 dB cutoff. (b) Extracted 3 dB bandwidth values as a function of metasurface width $W_m$.}\label{fig4}
\end{figure}

\section{Beamsplitting metasurface modulator}\label{sec6}

Expanding on the amplitude modulation capabilities of this platform, we demonstrate the general applicability of this nanoantenna platform as a design element for wavefront shaping metasurfaces. A key feature of these guided mode resonant nanoantennas is that each nanoantenna in the metasurface can have a different geometry for a tailored amplitude or phase response while still supporting high-Q resonances\cite{Lawrence2020d,Lin2023,Klopfer2023}. This allows for both near-field enhancement and far-field control of light. As such, each nanoantenna can be considered as a phase pixel. Figure 5a shows the simulated amplitude and phase response across the resonance for a design with $d$ = 80 nm, and nanoantenna spacing $w_p$ = 1400 nm. Across the resonance, we see a nearly $\pi/2$ phase shift. Arraying resonant (notched) and nonresonant (unnotched) antennas next to each other then leads to phase interference between antennas around the resonant wavelength. Figure 5b shows the electric field of one resonant and one nonresonant antenna next to each other in the metasurface supercell at the resonant wavelength of $\lambda$ = 1577.3 nm. This added design degree of freedom allows us to tailor the diffractive response of the metasurface. On the resonance, the antennas are out of phase with respect to each other, and light is preferentially transmitted to the $\pm$1st orders. Thus, this design acts as a beamsplitting metasurface. Figure 5c shows the diffracted intensity, with the 0th order decreased on the resonance and the $\pm$1st orders increased.

We fabricate these beamsplitting metasurfaces with the same procedures as the previous devices. Figure 5d shows a false color SEM of the fabricated beamsplitter metasurface. We experimentally characterize the diffracted intensity using Fourier plane imaging and spectroscopy, shown in figure 5e for a beamsplitter with device area of 200 $\times$ 200 $\mu$m. Off the resonance, the majority of light is sent to the 0th order, while on the resonance, light is split to the ±1st orders as well. We design this structure to have a relatively low Q of 780, as to maximize the amplitude contrast and resulting diffracted intensity to the ±1st orders. We see a maximum diffracted intensity of 12\% to the combined $\pm$1st orders on the resonance. With this beamsplitter metasurface, intensity modulated signals can be transmitted through each diffraction order, opening multiple channels for communication. Figure 5f shows the modulation amplitude measured in each diffraction order with $\pm$5 V at 100 MHz applied to the device, with the maximum change in transmittance in the 0th ($\pm$1st) orders being 0.12 (0.096)\%. The samples for this measurement were fabricated with 8 mm long electrode pads to compensate for the limitations of the optical setup. We attribute a lower modulation amplitude on these samples compared to the previous samples to a lower quality factor resonance as well as a nonideal voltage response from the long electrodes. 

We investigate the high frequency modulation response of these devices in figure 5g, sweeping the frequency of the applied voltage from 500 kHz - 1.6 GHz. We see an increase in the modulation bandwidth as we decrease the area of the device from 200 $\times$ 200 to 140 $\times$ 140 $\mu$m., measuring 3 dB bandwidth values of 430 MHz and 1.03 GHz respectively. Additionally, a resonant feature is present in these frequency sweeps around 1 GHz, likely arising from the periodicity of the electrodes potentially exciting an acoustic wave within the LNO. We confirm this by looking at the reflection coefficient ($S_{11}$) response of the electrodes using a vector network analyzer and see the same resonant feature in the purely electric response (see figure S12). Defining the cutoff frequency where the modulation response drops below -3 dB past this electrical resonance gives a bandwidth of 1.22 GHz, and further indicates that improvements in the electrode design, rather than the optical design, can further improve the response of these devices. 

\begin{figure}[h]
\centering
\includegraphics[width=\textwidth]{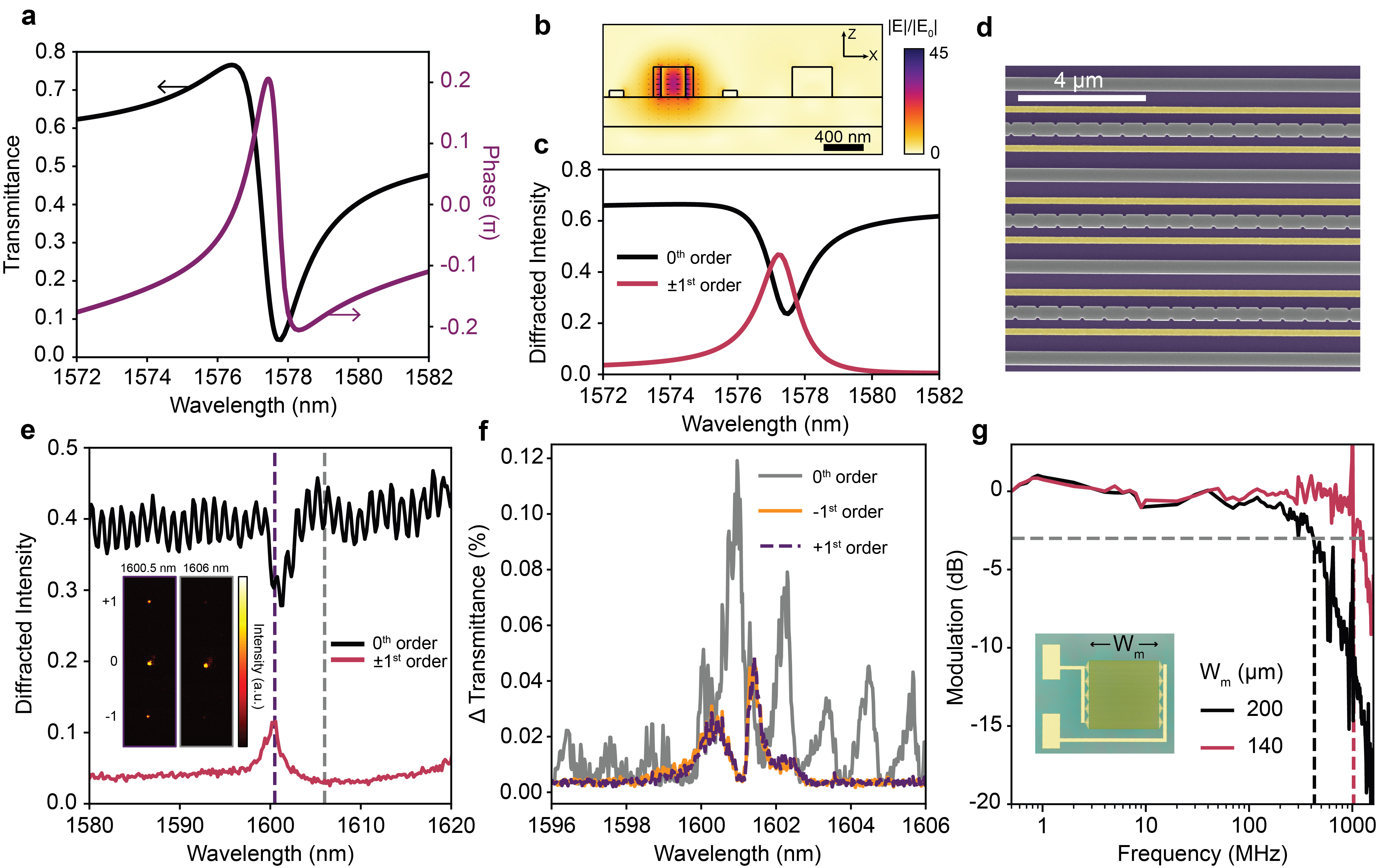}
\caption{(a) Simulated transmittance and phase response across the resonance in a metasurface with $d$ = 80 nm and $w_p$ = 1400 nm. (b) Simulated nearfield profile a single period from the high-Q beamsplitter metasurface comprising one resonant and one nonresonant nanoantenna at 1577.3 nm. (c) Simulated diffracted intensity for a high-Q beamsplitter metasurface. (d) SEM of fabricated beamsplitter metasurface. (e) Measured diffracted intensity from beamsplitter metasurface of 200 $\times$ 200 $\mu$m area. Inset shows Fourier plane imaging of diffracted beams on and off resonance. (f) Measured modulation amplitude for each diffraction order of the beamsplitter metasurface of 200 $\times$ 200 $\mu$m area. (g) Modulation bandwidth measurement from the beamsplitter metasurface, varying area of the metasurface device.}\label{fig5}
\end{figure}

\section{Conclusion}\label{sec7}

We have demonstrated that high frequency electro-optic modulation of individual nanoantennas can enable amplitude and wavefront modulation in a compact and efficient manner using a silicon on LNO metasurface platform. The high quality factor resonances (Q $>$ 1,000) from one-dimensionally confined guided modes in silicon nanoantennas enable modulation when voltage is applied to tune the refractive index of a LNO film interfaced with the nanoantennas. We demonstrate amplitude modulation from a resonant measurface, with peak modulation amplitude on the guided mode resonance of the metasurface. The transmittance modulation could be further optimized by engineering the Fano-resonant lineshape to maximize the amplitude contrast within the same linewidth\cite{Limonov2017b}. With the same design strategy, we develop a reconfigurable high-Q beamsplitter. This opens multiple channels for transmitting modulated signals, which we experimentally demonstrate to surpass 1 GHz in bandwidth, limited by the capacitance of the electrode structure. To our knowledge, this work presents the first LNO-based metasurface with a 3 dB bandwidth greater than 1 GHz, and the highest absolute transmittance modulation of electro-optic metasurfaces based on LNO or electro-optic polymers (see supplementary note 6). While we have focused on single frequency driving voltages for wavefront shaping, applying more complicated waveforms to the metasurface can control many degrees of freedom, such as in space-time metasurfaces\cite{Sisler2024}. With improvements in other superlative electro-optic materials performance, this heterogeneous materials platform can easily be integrated to demonstrate even more dramatic transmission contrast and modulation speed\cite{Benea-Chelmus2022,Zheng2024,Abel2019}. In addition to a large electro-optic response, LNO possesses large optical nonlinearities as well as photoelastic/piezoelectric responses. These properties provide other routes to wavefront shaping and modulation that can be enhanced by the nanoantenna design strategy presented here\cite{Jiang2020,Santiago-Cruz2021,Zhang2022}. Our resonant metasurface platform provides a general foundation for combining high frequency modulation with phase gradient enabled wavefront shaping, enabling a new suite of efficient and high speed reconfigurable optical elements with potential applications in ultrathin dynamic optics, free-space communications, and sensing.

\section{Methods}
\subsection{Computational design}

Electromagnetic and electrostatic simulations were performed using COMSOL Multiphysics. Metasurfaces were simulated with periodic boundary conditions in the x and y directions. An x-polarized plane wave illumination excites the structures from periodic ports, propagating in the z-direction. The incident port illuminates through a 3 $\mu$m LNO substrate region, and transmits through a 2 $\mu$m SiO$_2$ layer before reaching the Si-on-LNO nanoantenna. The transmission port is in the air region 5 $\mu$m above the nanoantenna layer. Electrical tuning simulations are conducted by solving for the electrostatic field and the resulting refractive index change in the thin film LNO layer. Full field electromagnetic simulations are then completed using the electrostatic refractive index profile at different voltage values. To calculate the fraction of the mode volume inside the LNO region, we use the conventional definition where $V_m$ is determined using the magnitude of the electric field ($E$) and the permittivity ($\epsilon$):

\begin{equation}
    V_m = \frac{\int\epsilon |E|^2 dV}{max(\epsilon |E|^2)}
\end{equation}

\subsection{Device fabrication}

First, commercially available silicon-on-insulator (UniversityWafer, 300 nm Si/300 nm SiO$_2$/725 $\mu$m Si substrate) and lithium niobate-on-insulator (NanoLN, 300 nm LNO/2 $\mu$m SiO$_2$/525 $\mu$m LNO substrate) pieces ($\sim$10 mm $\times$ 10 mm) were solvent cleaned in acetone, methanol, and isopropanol, and subsequently cleaned in a Piranha solution (9:1 H$_2$SO$_4$:H$_2$O$_2$). The chips were exposed to an O$_2$ plasma, then dipped in DI water for 30 seconds and dried before pressing the film surfaces together with tweezers to directly bond the films together at room temperature. Following an 8-16 hour anneal at 90\textdegree C, the substrate of the silicon-on-insulator is etched through a combination of dry etch (Plasma Therm Versaline LL ICP Deep Silicon Etcher, SF$_6$/C$_4$F$_8$ gases) and wet etch (6:1 Buffered Oxide Etch) steps. The Si metasurface pattern is written by exposing a negative tone resist. Hydrogen silsesquioxane (HSQ, XR-1571-006, DuPont) is baked at 80\textdegree C for 5 minutes before a conductive polymer (e-spacer, Resonac) is spun on to reduce charging during exposure. The resist is exposed using electron beam lithography (Raith EBPG 5200+), and developed in a salty aqueous developer (1\% NaOH, 4\% NaCl) for two minutes. The pattern is transferred into the Si layer using HBr and Cl dry etch chemistries in an inductively coupled plasma reactive ion etcher (Oxford III-V etcher).  Following etching, the sample is cleaned using a combination of HF and Piranha solution to remove redeposition and remove organic residue. The electrode layer is patterned in a bilayer PMMA resist (PMMA 495 A4, 950 A3) using electron beam lithography and developed in 1:3 MIBK:IPA for 90 seconds. 5 nm of Ti and 65 nm of Au is electron beam evaporated onto the sample (Kurt J. Lesker LAB18) and metal liftoff is completed during an 8 hour soak in n-methyl pyrrolidone (Remover PG) heated to 80\textdegree C.

\subsection{Electro-optic characterization}

A home-built transmission microscope setup is used to characterize the metasurface devices (see figures S10 and S11). Resonant spectra are characterized using a broadband supercontinuum laser source (NKT SuperK EXTREME). Electro-optic modulation measurements are done using a tunable laser source (Santec TSL 550). For wavelength spectra, amplitude modulation, and bandwidth measurements, transmitted light is collected with a 10x objective (Mitutoyo Plan Apo NIR). Wavelength spectra are recorded using a spectrometer (Princeton Instruments SPR-2300) and thermoelectrically cooled InGaAs detector (NiRvana, Princeton Instruments). For beamsplitting measurements, transmitted light is collected using a 50x objective (Olympus LCPlan N). Fourier plane imaging of the diffracted beams is recorded using a thermoelectrically cooled camera (NIT HiPe SenS 640 V-ST). Voltage is applied using a microwave probe (FormFactor ACP-250) connected to a signal generator (Rohde $\&$ Schwarz SMA100B). Wavelength dependent modulation measurements are recorded using a DC - 400 MHz bandwidth photodetector (Thorlabs APD430C/M), and frequency dependent modulation measurements are measured using a 300 kHz - 1.6 GHz bandwidth photodetector (Thorlabs APD 450C). The photodetectors are connected to a spectrum analyzer (Signal Hound BB60C). Modulation signals measured as a function of time were recorded using an oscilloscope (Agilent DSO-X 2024A). Background signals of the laser intensity were measured with the sample removed from the beam path to determine Transmittance and $\Delta$Transmittance values. 

The simulated and measured resonant spectral features were fit with the function:

\begin{equation}
    T = \left| \frac{1}{1+Fsin^2(n_skh_s)} \right| \left| a_r+a_ii+\frac{b}{f-f_0+i\gamma}\right|^2
\end{equation}

The first term accounts for the Fabry-Perot interference through the substrate of thickness $h_s$ and refractive index $n_s$. $k$ is the free-space wavevector ($2\pi/\lambda$) and $F$ accounts for the reflectivity of the air-substrate interfaces. The second term represents the superposition between a constant complex background, $a_r+a_ii$ and a Lorentzian resonance with resonant frequency $f_0$ and full-width at half-maximum $2\gamma$. The Q-factor of this resonance is then taken to be $Q = f_0/2\gamma$.

\backmatter

\bmhead{Acknowledgements}

The authors thank Mark Lawrence, Jason Herrmann, Feng Pan, Harsha Reddy, Stanley Lin, and Skyler Selvin for helpful discussions. The authors acknowledge support from the Office of Naval Research under the Multi University Research Initiative (MURI) program (award N00014-23-1-2567), which supported the salaries of S.D., C.-Y.C., and J.A.D.; and the U.S. Department of Energy Office of Science National Quantum Information Science Research Centers as part of the Q-NEXT center, which supported the device fabrication and salaries of S.D. and J.A.D. S.D. was also supported by the US Department of Defense through the National Defense Science and Engineering Graduate (NDSEG) Fellowship Program. H.B.B. acknowledges support from the HHMI Hanna H. Gray Fellowship. J.-H.S. acknowledges support from the Ministry of Education, Singapore, under the Academic Research Fund Tier 1 (FY2024). Work was performed in part at the Stanford Nanofabrication Facility (SNF) and the Stanford Nano Shared Facilities (SNSF), supported by the National Science Foundation (awards ECCS-1542152 and ECCS-2026822, respectively). 

\bmhead{Author Contributions}

S.D., E.K., D.B., and J.A.D. conceived the study. S.D. conducted the theory and numerical simulations under the guidance of D.B. and J.A.D. S.D., J.S., H.C.D., S.A., C.-Y.C., V.D., J.D., and B.O fabricated the devices. S.D. built the characterization setup with input from H.B.B., S.A., J.H., and J.-H.S. S.D. and J.S. collected experimental data. S.D. and J.S. analyzed the data under the guidance of M.L.B., D.B., and J.A.D. All authors contributed to the preparation of this manuscript. 

\bmhead{Declarations}
The authors declare no competing interests.

\bmhead{Additional information}
Supplementary Information is available for this paper. Correspondence and requests for materials should be addressed to S.D. (daglis@stanford.edu), D.B. (dbarton@northwestern.edu), and J.A.D. (jdionne@stanford.edu).

\bibliography{main}


\begin{thebibliography}{73}
\ifx \bisbn   \undefined \def \bisbn  #1{ISBN #1}\fi
\ifx \binits  \undefined \def \binits#1{#1}\fi
\ifx \bauthor  \undefined \def \bauthor#1{#1}\fi
\ifx \batitle  \undefined \def \batitle#1{#1}\fi
\ifx \bjtitle  \undefined \def \bjtitle#1{#1}\fi
\ifx \bvolume  \undefined \def \bvolume#1{\textbf{#1}}\fi
\ifx \byear  \undefined \def \byear#1{#1}\fi
\ifx \bissue  \undefined \def \bissue#1{#1}\fi
\ifx \bfpage  \undefined \def \bfpage#1{#1}\fi
\ifx \blpage  \undefined \def \blpage #1{#1}\fi
\ifx \burl  \undefined \def \burl#1{\textsf{#1}}\fi
\ifx \doiurl  \undefined \def \doiurl#1{\url{https://doi.org/#1}}\fi
\ifx \betal  \undefined \def \betal{\textit{et al.}}\fi
\ifx \binstitute  \undefined \def \binstitute#1{#1}\fi
\ifx \binstitutionaled  \undefined \def \binstitutionaled#1{#1}\fi
\ifx \bctitle  \undefined \def \bctitle#1{#1}\fi
\ifx \beditor  \undefined \def \beditor#1{#1}\fi
\ifx \bpublisher  \undefined \def \bpublisher#1{#1}\fi
\ifx \bbtitle  \undefined \def \bbtitle#1{#1}\fi
\ifx \bedition  \undefined \def \bedition#1{#1}\fi
\ifx \bseriesno  \undefined \def \bseriesno#1{#1}\fi
\ifx \blocation  \undefined \def \blocation#1{#1}\fi
\ifx \bsertitle  \undefined \def \bsertitle#1{#1}\fi
\ifx \bsnm \undefined \def \bsnm#1{#1}\fi
\ifx \bsuffix \undefined \def \bsuffix#1{#1}\fi
\ifx \bparticle \undefined \def \bparticle#1{#1}\fi
\ifx \barticle \undefined \def \barticle#1{#1}\fi
\bibcommenthead
\ifx \bconfdate \undefined \def \bconfdate #1{#1}\fi
\ifx \botherref \undefined \def \botherref #1{#1}\fi
\ifx \url \undefined \def \url#1{\textsf{#1}}\fi
\ifx \bchapter \undefined \def \bchapter#1{#1}\fi
\ifx \bbook \undefined \def \bbook#1{#1}\fi
\ifx \bcomment \undefined \def \bcomment#1{#1}\fi
\ifx \oauthor \undefined \def \oauthor#1{#1}\fi
\ifx \citeauthoryear \undefined \def \citeauthoryear#1{#1}\fi
\ifx \endbibitem  \undefined \def \endbibitem {}\fi
\ifx \bconflocation  \undefined \def \bconflocation#1{#1}\fi
\ifx \arxivurl  \undefined \def \arxivurl#1{\textsf{#1}}\fi
\csname PreBibitemsHook\endcsname

\bibitem[\protect\citeauthoryear{Liu et~al.}{2015}]{Liu2015}
\begin{barticle}
\bauthor{\bsnm{Liu}, \binits{K.}},
\bauthor{\bsnm{Ye}, \binits{C.R.}},
\bauthor{\bsnm{Khan}, \binits{S.}},
\bauthor{\bsnm{Sorger}, \binits{V.J.}}:
\batitle{Review and perspective on ultrafast wavelength-size electro-optic modulators}.
\bjtitle{Laser \& Photonics Reviews}
\bvolume{9},
\bfpage{172}--\blpage{194}
(\byear{2015})
\doiurl{10.1002/lpor.201400219}
\end{barticle}
\endbibitem

\bibitem[\protect\citeauthoryear{Sinatkas et~al.}{2021}]{Sinatkas2021}
\begin{botherref}
\oauthor{\bsnm{Sinatkas}, \binits{G.}},
\oauthor{\bsnm{Christopoulos}, \binits{T.}},
\oauthor{\bsnm{Tsilipakos}, \binits{O.}},
\oauthor{\bsnm{Kriezis}, \binits{E.E.}}:
Electro-optic modulation in integrated photonics.
Journal of Applied Physics
\textbf{130}
(2021)
\doiurl{10.1063/5.0048712}
\end{botherref}
\endbibitem

\bibitem[\protect\citeauthoryear{Kim et~al.}{2021}]{Kim2021}
\begin{barticle}
\bauthor{\bsnm{Kim}, \binits{I.}},
\bauthor{\bsnm{Martins}, \binits{R.J.}},
\bauthor{\bsnm{Jang}, \binits{J.}},
\bauthor{\bsnm{Badloe}, \binits{T.}},
\bauthor{\bsnm{Khadir}, \binits{S.}},
\bauthor{\bsnm{Jung}, \binits{H.-Y.}},
\bauthor{\bsnm{Kim}, \binits{H.}},
\bauthor{\bsnm{Kim}, \binits{J.}},
\bauthor{\bsnm{Genevet}, \binits{P.}},
\bauthor{\bsnm{Rho}, \binits{J.}}:
\batitle{Nanophotonics for light detection and ranging technology}.
\bjtitle{Nature Nanotechnology}
\bvolume{16},
\bfpage{508}--\blpage{524}
(\byear{2021})
\doiurl{10.1038/s41565-021-00895-3}
\end{barticle}
\endbibitem

\bibitem[\protect\citeauthoryear{Abumarshoud et~al.}{2022}]{Abumarshoud2022}
\begin{barticle}
\bauthor{\bsnm{Abumarshoud}, \binits{H.}},
\bauthor{\bsnm{Mohjazi}, \binits{L.}},
\bauthor{\bsnm{Dobre}, \binits{O.A.}},
\bauthor{\bsnm{Renzo}, \binits{M.D.}},
\bauthor{\bsnm{Imran}, \binits{M.A.}},
\bauthor{\bsnm{Haas}, \binits{H.}}:
\batitle{Lifi through reconfigurable intelligent surfaces: A new frontier for 6g?}
\bjtitle{IEEE Vehicular Technology Magazine}
\bvolume{17},
\bfpage{37}--\blpage{46}
(\byear{2022})
\doiurl{10.1109/MVT.2021.3121647}
\end{barticle}
\endbibitem

\bibitem[\protect\citeauthoryear{Thureja et~al.}{2022}]{Thureja2022}
\begin{barticle}
\bauthor{\bsnm{Thureja}, \binits{P.}},
\bauthor{\bsnm{Sokhoyan}, \binits{R.}},
\bauthor{\bsnm{Hail}, \binits{C.U.}},
\bauthor{\bsnm{Sisler}, \binits{J.}},
\bauthor{\bsnm{Foley}, \binits{M.}},
\bauthor{\bsnm{Grajower}, \binits{M.Y.}},
\bauthor{\bsnm{Atwater}, \binits{H.A.}}:
\batitle{Toward a universal metasurface for optical imaging, communication, and computation}.
\bjtitle{Nanophotonics}
\bvolume{11},
\bfpage{3745}--\blpage{3768}
(\byear{2022})
\doiurl{10.1515/nanoph-2022-0155}
\end{barticle}
\endbibitem

\bibitem[\protect\citeauthoryear{Gopakumar et~al.}{2024}]{Gopakumar2024}
\begin{barticle}
\bauthor{\bsnm{Gopakumar}, \binits{M.}},
\bauthor{\bsnm{Lee}, \binits{G.-Y.}},
\bauthor{\bsnm{Choi}, \binits{S.}},
\bauthor{\bsnm{Chao}, \binits{B.}},
\bauthor{\bsnm{Peng}, \binits{Y.}},
\bauthor{\bsnm{Kim}, \binits{J.}},
\bauthor{\bsnm{Wetzstein}, \binits{G.}}:
\batitle{Full-colour 3d holographic augmented-reality displays with metasurface waveguides}.
\bjtitle{Nature}
\bvolume{629},
\bfpage{791}--\blpage{797}
(\byear{2024})
\doiurl{10.1038/s41586-024-07386-0}
\end{barticle}
\endbibitem

\bibitem[\protect\citeauthoryear{Song et~al.}{2021}]{Song2021}
\begin{barticle}
\bauthor{\bsnm{Song}, \binits{J.-H.}},
\bauthor{\bsnm{Groep}, \binits{J.}},
\bauthor{\bsnm{Kim}, \binits{S.J.}},
\bauthor{\bsnm{Brongersma}, \binits{M.L.}}:
\batitle{Non-local metasurfaces for spectrally decoupled wavefront manipulation and eye tracking}.
\bjtitle{Nature Nanotechnology}
\bvolume{16},
\bfpage{1224}--\blpage{1230}
(\byear{2021})
\doiurl{10.1038/s41565-021-00967-4}
\end{barticle}
\endbibitem

\bibitem[\protect\citeauthoryear{Huang et~al.}{2018}]{Huang2018}
\begin{barticle}
\bauthor{\bsnm{Huang}, \binits{L.}},
\bauthor{\bsnm{Zhang}, \binits{S.}},
\bauthor{\bsnm{Zentgraf}, \binits{T.}}:
\batitle{Metasurface holography: from fundamentals to applications}.
\bjtitle{Nanophotonics}
\bvolume{7},
\bfpage{1169}--\blpage{1190}
(\byear{2018})
\doiurl{10.1515/nanoph-2017-0118}
\end{barticle}
\endbibitem

\bibitem[\protect\citeauthoryear{Xiong et~al.}{2023}]{Xiong2023}
\begin{barticle}
\bauthor{\bsnm{Xiong}, \binits{B.}},
\bauthor{\bsnm{Liu}, \binits{Y.}},
\bauthor{\bsnm{Xu}, \binits{Y.}},
\bauthor{\bsnm{Deng}, \binits{L.}},
\bauthor{\bsnm{Chen}, \binits{C.-W.}},
\bauthor{\bsnm{Wang}, \binits{J.-N.}},
\bauthor{\bsnm{Peng}, \binits{R.}},
\bauthor{\bsnm{Lai}, \binits{Y.}},
\bauthor{\bsnm{Liu}, \binits{Y.}},
\bauthor{\bsnm{Wang}, \binits{M.}}:
\batitle{Breaking the limitation of polarization multiplexing in optical metasurfaces with engineered noise}.
\bjtitle{Science}
\bvolume{379},
\bfpage{294}--\blpage{299}
(\byear{2023})
\doiurl{10.1126/science.ade5140}
\end{barticle}
\endbibitem

\bibitem[\protect\citeauthoryear{Solntsev et~al.}{2021}]{Solntsev2021}
\begin{barticle}
\bauthor{\bsnm{Solntsev}, \binits{A.S.}},
\bauthor{\bsnm{Agarwal}, \binits{G.S.}},
\bauthor{\bsnm{Kivshar}, \binits{Y.S.}}:
\batitle{Metasurfaces for quantum photonics}.
\bjtitle{Nature Photonics}
\bvolume{15},
\bfpage{327}--\blpage{336}
(\byear{2021})
\doiurl{10.1038/s41566-021-00793-z}
\end{barticle}
\endbibitem

\bibitem[\protect\citeauthoryear{Kuznetsov et~al.}{2024}]{Kuznetsov2024}
\begin{barticle}
\bauthor{\bsnm{Kuznetsov}, \binits{A.I.}},
\bauthor{\bsnm{Brongersma}, \binits{M.L.}},
\bauthor{\bsnm{Yao}, \binits{J.}},
\bauthor{\bsnm{Chen}, \binits{M.K.}},
\bauthor{\bsnm{Levy}, \binits{U.}},
\bauthor{\bsnm{Tsai}, \binits{D.P.}},
\bauthor{\bsnm{Zheludev}, \binits{N.I.}},
\bauthor{\bsnm{Faraon}, \binits{A.}},
\bauthor{\bsnm{Arbabi}, \binits{A.}},
\bauthor{\bsnm{Yu}, \binits{N.}},
\bauthor{\bsnm{Chanda}, \binits{D.}},
\bauthor{\bsnm{Crozier}, \binits{K.B.}},
\bauthor{\bsnm{Kildishev}, \binits{A.V.}},
\bauthor{\bsnm{Wang}, \binits{H.}},
\bauthor{\bsnm{Yang}, \binits{J.K.W.}},
\bauthor{\bsnm{Valentine}, \binits{J.G.}},
\bauthor{\bsnm{Genevet}, \binits{P.}},
\bauthor{\bsnm{Fan}, \binits{J.A.}},
\bauthor{\bsnm{Miller}, \binits{O.D.}},
\bauthor{\bsnm{Majumdar}, \binits{A.}},
\bauthor{\bsnm{Fröch}, \binits{J.E.}},
\bauthor{\bsnm{Brady}, \binits{D.}},
\bauthor{\bsnm{Heide}, \binits{F.}},
\bauthor{\bsnm{Veeraraghavan}, \binits{A.}},
\bauthor{\bsnm{Engheta}, \binits{N.}},
\bauthor{\bsnm{Alù}, \binits{A.}},
\bauthor{\bsnm{Polman}, \binits{A.}},
\bauthor{\bsnm{Atwater}, \binits{H.A.}},
\bauthor{\bsnm{Thureja}, \binits{P.}},
\bauthor{\bsnm{Paniagua-Dominguez}, \binits{R.}},
\bauthor{\bsnm{Ha}, \binits{S.T.}},
\bauthor{\bsnm{Barreda}, \binits{A.I.}},
\bauthor{\bsnm{Schuller}, \binits{J.A.}},
\bauthor{\bsnm{Staude}, \binits{I.}},
\bauthor{\bsnm{Grinblat}, \binits{G.}},
\bauthor{\bsnm{Kivshar}, \binits{Y.}},
\bauthor{\bsnm{Peana}, \binits{S.}},
\bauthor{\bsnm{Yelin}, \binits{S.F.}},
\bauthor{\bsnm{Senichev}, \binits{A.}},
\bauthor{\bsnm{Shalaev}, \binits{V.M.}},
\bauthor{\bsnm{Saha}, \binits{S.}},
\bauthor{\bsnm{Boltasseva}, \binits{A.}},
\bauthor{\bsnm{Rho}, \binits{J.}},
\bauthor{\bsnm{Oh}, \binits{D.K.}},
\bauthor{\bsnm{Kim}, \binits{J.}},
\bauthor{\bsnm{Park}, \binits{J.}},
\bauthor{\bsnm{Devlin}, \binits{R.}},
\bauthor{\bsnm{Pala}, \binits{R.A.}}:
\batitle{Roadmap for optical metasurfaces}.
\bjtitle{ACS Photonics}
\bvolume{11},
\bfpage{816}--\blpage{865}
(\byear{2024})
\doiurl{10.1021/acsphotonics.3c00457}
\end{barticle}
\endbibitem

\bibitem[\protect\citeauthoryear{Jung et~al.}{2024}]{Jung2024}
\begin{botherref}
\oauthor{\bsnm{Jung}, \binits{C.}},
\oauthor{\bsnm{Lee}, \binits{E.}},
\oauthor{\bsnm{Rho}, \binits{J.}}:
The rise of electrically tunable metasurfaces.
Science Advances
\textbf{10}
(2024)
\doiurl{10.1126/sciadv.ado8964}
\end{botherref}
\endbibitem

\bibitem[\protect\citeauthoryear{Shaltout et~al.}{2019}]{Shaltout2019f}
\begin{botherref}
\oauthor{\bsnm{Shaltout}, \binits{A.M.}},
\oauthor{\bsnm{Shalaev}, \binits{V.M.}},
\oauthor{\bsnm{Brongersma}, \binits{M.L.}}:
Spatiotemporal light control with active metasurfaces.
Science
\textbf{364}
(2019)
\doiurl{10.1126/science.aat3100}
\end{botherref}
\endbibitem

\bibitem[\protect\citeauthoryear{Ha et~al.}{2024}]{Ha2024}
\begin{botherref}
\oauthor{\bsnm{Ha}, \binits{S.T.}},
\oauthor{\bsnm{Li}, \binits{Q.}},
\oauthor{\bsnm{Yang}, \binits{J.K.W.}},
\oauthor{\bsnm{Demir}, \binits{H.V.}},
\oauthor{\bsnm{Brongersma}, \binits{M.L.}},
\oauthor{\bsnm{Kuznetsov}, \binits{A.I.}}:
Optoelectronic metadevices.
Science
\textbf{386}
(2024)
\doiurl{10.1126/science.adm7442}
\end{botherref}
\endbibitem

\bibitem[\protect\citeauthoryear{Li et~al.}{2019}]{Li2019s}
\begin{barticle}
\bauthor{\bsnm{Li}, \binits{S.-Q.}},
\bauthor{\bsnm{Xu}, \binits{X.}},
\bauthor{\bsnm{Veetil}, \binits{R.M.}},
\bauthor{\bsnm{Valuckas}, \binits{V.}},
\bauthor{\bsnm{Paniagua-Domínguez}, \binits{R.}},
\bauthor{\bsnm{Kuznetsov}, \binits{A.I.}}:
\batitle{Phase-only transmissive spatial light modulator based on tunable dielectric metasurface}.
\bjtitle{Science}
\bvolume{364},
\bfpage{1087}--\blpage{1090}
(\byear{2019})
\doiurl{10.1126/science.aaw6747}
\end{barticle}
\endbibitem

\bibitem[\protect\citeauthoryear{Moitra et~al.}{2023}]{Moitra2023}
\begin{barticle}
\bauthor{\bsnm{Moitra}, \binits{P.}},
\bauthor{\bsnm{Xu}, \binits{X.}},
\bauthor{\bsnm{Veetil}, \binits{R.M.}},
\bauthor{\bsnm{Liang}, \binits{X.}},
\bauthor{\bsnm{Mass}, \binits{T.W.W.}},
\bauthor{\bsnm{Kuznetsov}, \binits{A.I.}},
\bauthor{\bsnm{Paniagua-Domínguez}, \binits{R.}}:
\batitle{Electrically tunable reflective metasurfaces with continuous and full-phase modulation for high-efficiency wavefront control at visible frequencies}.
\bjtitle{ACS Nano}
\bvolume{17},
\bfpage{16952}--\blpage{16959}
(\byear{2023})
\doiurl{10.1021/acsnano.3c04071}
\end{barticle}
\endbibitem

\bibitem[\protect\citeauthoryear{Cai et~al.}{2024}]{Cai2024}
\begin{barticle}
\bauthor{\bsnm{Cai}, \binits{G.}},
\bauthor{\bsnm{Li}, \binits{Y.}},
\bauthor{\bsnm{Zhang}, \binits{Y.}},
\bauthor{\bsnm{Jiang}, \binits{X.}},
\bauthor{\bsnm{Chen}, \binits{Y.}},
\bauthor{\bsnm{Qu}, \binits{G.}},
\bauthor{\bsnm{Zhang}, \binits{X.}},
\bauthor{\bsnm{Xiao}, \binits{S.}},
\bauthor{\bsnm{Han}, \binits{J.}},
\bauthor{\bsnm{Yu}, \binits{S.}},
\bauthor{\bsnm{Kivshar}, \binits{Y.}},
\bauthor{\bsnm{Song}, \binits{Q.}}:
\batitle{Compact angle-resolved metasurface spectrometer}.
\bjtitle{Nature Materials}
\bvolume{23},
\bfpage{71}--\blpage{78}
(\byear{2024})
\doiurl{10.1038/s41563-023-01710-1}
\end{barticle}
\endbibitem

\bibitem[\protect\citeauthoryear{Zhang et~al.}{2021}]{Zhang2021}
\begin{barticle}
\bauthor{\bsnm{Zhang}, \binits{Y.}},
\bauthor{\bsnm{Fowler}, \binits{C.}},
\bauthor{\bsnm{Liang}, \binits{J.}},
\bauthor{\bsnm{Azhar}, \binits{B.}},
\bauthor{\bsnm{Shalaginov}, \binits{M.Y.}},
\bauthor{\bsnm{Deckoff-Jones}, \binits{S.}},
\bauthor{\bsnm{An}, \binits{S.}},
\bauthor{\bsnm{Chou}, \binits{J.B.}},
\bauthor{\bsnm{Roberts}, \binits{C.M.}},
\bauthor{\bsnm{Liberman}, \binits{V.}},
\bauthor{\bsnm{Kang}, \binits{M.}},
\bauthor{\bsnm{R{\'{i}}os}, \binits{C.}},
\bauthor{\bsnm{Richardson}, \binits{K.A.}},
\bauthor{\bsnm{Rivero-Baleine}, \binits{C.}},
\bauthor{\bsnm{Gu}, \binits{T.}},
\bauthor{\bsnm{Zhang}, \binits{H.}},
\bauthor{\bsnm{Hu}, \binits{J.}}:
\batitle{{Electrically reconfigurable non-volatile metasurface using low-loss optical phase-change material}}.
\bjtitle{Nature Nanotechnology}
\bvolume{16}(\bissue{6}),
\bfpage{661}--\blpage{666}
(\byear{2021})
\doiurl{10.1038/s41565-021-00881-9}
\end{barticle}
\endbibitem

\bibitem[\protect\citeauthoryear{Wang et~al.}{2021}]{Wang2020a}
\begin{barticle}
\bauthor{\bsnm{Wang}, \binits{Y.}},
\bauthor{\bsnm{Landreman}, \binits{P.}},
\bauthor{\bsnm{Schoen}, \binits{D.}},
\bauthor{\bsnm{Okabe}, \binits{K.}},
\bauthor{\bsnm{Marshall}, \binits{A.}},
\bauthor{\bsnm{Celano}, \binits{U.}},
\bauthor{\bsnm{Wong}, \binits{H.-S.P.}},
\bauthor{\bsnm{Park}, \binits{J.}},
\bauthor{\bsnm{Brongersma}, \binits{M.L.}}:
\batitle{Electrical tuning of phase-change antennas and metasurfaces}.
\bjtitle{Nature Nanotechnology}
\bvolume{16},
\bfpage{667}--\blpage{672}
(\byear{2021})
\doiurl{10.1038/s41565-021-00882-8}
\end{barticle}
\endbibitem

\bibitem[\protect\citeauthoryear{Abdollahramezani et~al.}{2022}]{Abdollahramezani2022}
\begin{barticle}
\bauthor{\bsnm{Abdollahramezani}, \binits{S.}},
\bauthor{\bsnm{Hemmatyar}, \binits{O.}},
\bauthor{\bsnm{Taghinejad}, \binits{M.}},
\bauthor{\bsnm{Taghinejad}, \binits{H.}},
\bauthor{\bsnm{Krasnok}, \binits{A.}},
\bauthor{\bsnm{Eftekhar}, \binits{A.A.}},
\bauthor{\bsnm{Teichrib}, \binits{C.}},
\bauthor{\bsnm{Deshmukh}, \binits{S.}},
\bauthor{\bsnm{El-Sayed}, \binits{M.A.}},
\bauthor{\bsnm{Pop}, \binits{E.}},
\bauthor{\bsnm{Wuttig}, \binits{M.}},
\bauthor{\bsnm{Alù}, \binits{A.}},
\bauthor{\bsnm{Cai}, \binits{W.}},
\bauthor{\bsnm{Adibi}, \binits{A.}}:
\batitle{Electrically driven reprogrammable phase-change metasurface reaching 80\% efficiency}.
\bjtitle{Nature Communications}
\bvolume{13},
\bfpage{1696}
(\byear{2022})
\doiurl{10.1038/s41467-022-29374-6}
\end{barticle}
\endbibitem

\bibitem[\protect\citeauthoryear{Fang et~al.}{2024}]{Fang2024}
\begin{barticle}
\bauthor{\bsnm{Fang}, \binits{Z.}},
\bauthor{\bsnm{Chen}, \binits{R.}},
\bauthor{\bsnm{Fröch}, \binits{J.E.}},
\bauthor{\bsnm{Tanguy}, \binits{Q.A.A.}},
\bauthor{\bsnm{Khan}, \binits{A.I.}},
\bauthor{\bsnm{Wu}, \binits{X.}},
\bauthor{\bsnm{Tara}, \binits{V.}},
\bauthor{\bsnm{Manna}, \binits{A.}},
\bauthor{\bsnm{Sharp}, \binits{D.}},
\bauthor{\bsnm{Munley}, \binits{C.}},
\bauthor{\bsnm{Miller}, \binits{F.}},
\bauthor{\bsnm{Zhao}, \binits{Y.}},
\bauthor{\bsnm{Geiger}, \binits{S.}},
\bauthor{\bsnm{Böhringer}, \binits{K.F.}},
\bauthor{\bsnm{Reynolds}, \binits{M.S.}},
\bauthor{\bsnm{Pop}, \binits{E.}},
\bauthor{\bsnm{Majumdar}, \binits{A.}}:
\batitle{Nonvolatile phase-only transmissive spatial light modulator with electrical addressability of individual pixels}.
\bjtitle{ACS Nano}
\bvolume{18},
\bfpage{11245}--\blpage{11256}
(\byear{2024})
\doiurl{10.1021/acsnano.4c00340}
\end{barticle}
\endbibitem

\bibitem[\protect\citeauthoryear{Huang et~al.}{2016}]{Huang2016d}
\begin{barticle}
\bauthor{\bsnm{Huang}, \binits{Y.-W.}},
\bauthor{\bsnm{Lee}, \binits{H.W.H.}},
\bauthor{\bsnm{Sokhoyan}, \binits{R.}},
\bauthor{\bsnm{Pala}, \binits{R.A.}},
\bauthor{\bsnm{Thyagarajan}, \binits{K.}},
\bauthor{\bsnm{Han}, \binits{S.}},
\bauthor{\bsnm{Tsai}, \binits{D.P.}},
\bauthor{\bsnm{Atwater}, \binits{H.A.}}:
\batitle{Gate-tunable conducting oxide metasurfaces}.
\bjtitle{Nano Letters}
\bvolume{16},
\bfpage{5319}--\blpage{5325}
(\byear{2016})
\doiurl{10.1021/acs.nanolett.6b00555}
\end{barticle}
\endbibitem

\bibitem[\protect\citeauthoryear{Shirmanesh et~al.}{2020}]{Shirmanesh2020d}
\begin{barticle}
\bauthor{\bsnm{Shirmanesh}, \binits{G.K.}},
\bauthor{\bsnm{Sokhoyan}, \binits{R.}},
\bauthor{\bsnm{Wu}, \binits{P.C.}},
\bauthor{\bsnm{Atwater}, \binits{H.A.}}:
\batitle{Electro-optically tunable multifunctional metasurfaces}.
\bjtitle{ACS Nano}
\bvolume{14},
\bfpage{6912}--\blpage{6920}
(\byear{2020})
\doiurl{10.1021/acsnano.0c01269}
\end{barticle}
\endbibitem

\bibitem[\protect\citeauthoryear{Park et~al.}{2021}]{Park2020e}
\begin{barticle}
\bauthor{\bsnm{Park}, \binits{J.}},
\bauthor{\bsnm{Jeong}, \binits{B.G.}},
\bauthor{\bsnm{Kim}, \binits{S.I.}},
\bauthor{\bsnm{Lee}, \binits{D.}},
\bauthor{\bsnm{Kim}, \binits{J.}},
\bauthor{\bsnm{Shin}, \binits{C.}},
\bauthor{\bsnm{Lee}, \binits{C.B.}},
\bauthor{\bsnm{Otsuka}, \binits{T.}},
\bauthor{\bsnm{Kyoung}, \binits{J.}},
\bauthor{\bsnm{Kim}, \binits{S.}},
\bauthor{\bsnm{Yang}, \binits{K.-Y.}},
\bauthor{\bsnm{Park}, \binits{Y.-Y.}},
\bauthor{\bsnm{Lee}, \binits{J.}},
\bauthor{\bsnm{Hwang}, \binits{I.}},
\bauthor{\bsnm{Jang}, \binits{J.}},
\bauthor{\bsnm{Song}, \binits{S.H.}},
\bauthor{\bsnm{Brongersma}, \binits{M.L.}},
\bauthor{\bsnm{Ha}, \binits{K.}},
\bauthor{\bsnm{Hwang}, \binits{S.-W.}},
\bauthor{\bsnm{Choo}, \binits{H.}},
\bauthor{\bsnm{Choi}, \binits{B.L.}}:
\batitle{All-solid-state spatial light modulator with independent phase and amplitude control for three-dimensional lidar applications}.
\bjtitle{Nature Nanotechnology}
\bvolume{16},
\bfpage{69}--\blpage{76}
(\byear{2021})
\doiurl{10.1038/s41565-020-00787-y}
\end{barticle}
\endbibitem

\bibitem[\protect\citeauthoryear{Kim and Brongersma}{2017}]{Kim2017}
\begin{barticle}
\bauthor{\bsnm{Kim}, \binits{S.J.}},
\bauthor{\bsnm{Brongersma}, \binits{M.L.}}:
\batitle{Active flat optics using a guided mode resonance}.
\bjtitle{Optics Letters}
\bvolume{42},
\bfpage{5}
(\byear{2017})
\doiurl{10.1364/OL.42.000005}
\end{barticle}
\endbibitem

\bibitem[\protect\citeauthoryear{Weis and Gaylord}{1985}]{Weis1985c}
\begin{barticle}
\bauthor{\bsnm{Weis}, \binits{R.S.}},
\bauthor{\bsnm{Gaylord}, \binits{T.K.}}:
\batitle{Lithium niobate: Summary of physical properties and crystal structure}.
\bjtitle{Applied Physics A Solids and Surfaces}
\bvolume{37},
\bfpage{191}--\blpage{203}
(\byear{1985})
\doiurl{10.1007/BF00614817}
\end{barticle}
\endbibitem

\bibitem[\protect\citeauthoryear{Boes et~al.}{2023}]{Boes2023}
\begin{botherref}
\oauthor{\bsnm{Boes}, \binits{A.}},
\oauthor{\bsnm{Chang}, \binits{L.}},
\oauthor{\bsnm{Langrock}, \binits{C.}},
\oauthor{\bsnm{Yu}, \binits{M.}},
\oauthor{\bsnm{Zhang}, \binits{M.}},
\oauthor{\bsnm{Lin}, \binits{Q.}},
\oauthor{\bsnm{Lončar}, \binits{M.}},
\oauthor{\bsnm{Fejer}, \binits{M.}},
\oauthor{\bsnm{Bowers}, \binits{J.}},
\oauthor{\bsnm{Mitchell}, \binits{A.}}:
Lithium niobate photonics: Unlocking the electromagnetic spectrum.
Science
\textbf{379}
(2023)
\doiurl{10.1126/science.abj4396}
\end{botherref}
\endbibitem

\bibitem[\protect\citeauthoryear{Wang et~al.}{2018}]{Wang2018i}
\begin{barticle}
\bauthor{\bsnm{Wang}, \binits{C.}},
\bauthor{\bsnm{Zhang}, \binits{M.}},
\bauthor{\bsnm{Chen}, \binits{X.}},
\bauthor{\bsnm{Bertrand}, \binits{M.}},
\bauthor{\bsnm{Shams-Ansari}, \binits{A.}},
\bauthor{\bsnm{Chandrasekhar}, \binits{S.}},
\bauthor{\bsnm{Winzer}, \binits{P.}},
\bauthor{\bsnm{Lončar}, \binits{M.}}:
\batitle{Integrated lithium niobate electro-optic modulators operating at cmos-compatible voltages}.
\bjtitle{Nature}
\bvolume{562},
\bfpage{101}--\blpage{104}
(\byear{2018})
\doiurl{10.1038/s41586-018-0551-y}
\end{barticle}
\endbibitem

\bibitem[\protect\citeauthoryear{Gu et~al.}{2023}]{Gu2023}
\begin{barticle}
\bauthor{\bsnm{Gu}, \binits{T.}},
\bauthor{\bsnm{Kim}, \binits{H.J.}},
\bauthor{\bsnm{Rivero-Baleine}, \binits{C.}},
\bauthor{\bsnm{Hu}, \binits{J.}}:
\batitle{Reconfigurable metasurfaces towards commercial success}.
\bjtitle{Nature Photonics}
\bvolume{17},
\bfpage{48}--\blpage{58}
(\byear{2023})
\doiurl{10.1038/s41566-022-01099-4}
\end{barticle}
\endbibitem

\bibitem[\protect\citeauthoryear{Benea-Chelmus et~al.}{2022}]{Benea-Chelmus2022}
\begin{barticle}
\bauthor{\bsnm{Benea-Chelmus}, \binits{I.-C.}},
\bauthor{\bsnm{Mason}, \binits{S.}},
\bauthor{\bsnm{Meretska}, \binits{M.L.}},
\bauthor{\bsnm{Elder}, \binits{D.L.}},
\bauthor{\bsnm{Kazakov}, \binits{D.}},
\bauthor{\bsnm{Shams-Ansari}, \binits{A.}},
\bauthor{\bsnm{Dalton}, \binits{L.R.}},
\bauthor{\bsnm{Capasso}, \binits{F.}}:
\batitle{Gigahertz free-space electro-optic modulators based on mie resonances}.
\bjtitle{Nature Communications}
\bvolume{13},
\bfpage{3170}
(\byear{2022})
\doiurl{10.1038/s41467-022-30451-z}
\end{barticle}
\endbibitem

\bibitem[\protect\citeauthoryear{Zheng et~al.}{2024}]{Zheng2024}
\begin{barticle}
\bauthor{\bsnm{Zheng}, \binits{T.}},
\bauthor{\bsnm{Gu}, \binits{Y.}},
\bauthor{\bsnm{Kwon}, \binits{H.}},
\bauthor{\bsnm{Roberts}, \binits{G.}},
\bauthor{\bsnm{Faraon}, \binits{A.}}:
\batitle{Dynamic light manipulation via silicon-organic slot metasurfaces}.
\bjtitle{Nature Communications}
\bvolume{15},
\bfpage{1557}
(\byear{2024})
\doiurl{10.1038/s41467-024-45544-0}
\end{barticle}
\endbibitem

\bibitem[\protect\citeauthoryear{Lin et~al.}{2025}]{Lin2025}
\begin{botherref}
\oauthor{\bsnm{Lin}, \binits{S.}},
\oauthor{\bsnm{Chen}, \binits{Y.}},
\oauthor{\bsnm{Hwang}, \binits{T.}},
\oauthor{\bsnm{Upadhyay}, \binits{A.}},
\oauthor{\bsnm{Rady}, \binits{R.}},
\oauthor{\bsnm{Dolt}, \binits{D.}},
\oauthor{\bsnm{Palermo}, \binits{S.}},
\oauthor{\bsnm{Entesari}, \binits{K.}},
\oauthor{\bsnm{Madsen}, \binits{C.}},
\oauthor{\bsnm{Wong}, \binits{Z.J.}},
\oauthor{\bsnm{Lan}, \binits{S.}}:
Gigahertz directional light modulation with electro-optic metasurfaces.
arXiv preprint arXiv:2501.06102
(2025)
\end{botherref}
\endbibitem

\bibitem[\protect\citeauthoryear{Weigand et~al.}{2024}]{Weigand2024}
\begin{barticle}
\bauthor{\bsnm{Weigand}, \binits{H.C.}},
\bauthor{\bsnm{Talts}},
\bauthor{\bsnm{Vieli}, \binits{A.-L.}},
\bauthor{\bsnm{Vogler-Neuling}, \binits{V.V.}},
\bauthor{\bsnm{Nardi}, \binits{A.}},
\bauthor{\bsnm{Grange}, \binits{R.}}:
\batitle{Nanoimprinting solution-derived barium titanate for electro-optic metasurfaces}.
\bjtitle{Nano Letters}
\bvolume{24},
\bfpage{5536}--\blpage{5542}
(\byear{2024})
\doiurl{10.1021/acs.nanolett.4c00711}
\end{barticle}
\endbibitem

\bibitem[\protect\citeauthoryear{Abel et~al.}{2019}]{Abel2019}
\begin{barticle}
\bauthor{\bsnm{Abel}, \binits{S.}},
\bauthor{\bsnm{Eltes}, \binits{F.}},
\bauthor{\bsnm{Ortmann}, \binits{J.E.}},
\bauthor{\bsnm{Messner}, \binits{A.}},
\bauthor{\bsnm{Castera}, \binits{P.}},
\bauthor{\bsnm{Wagner}, \binits{T.}},
\bauthor{\bsnm{Urbonas}, \binits{D.}},
\bauthor{\bsnm{Rosa}, \binits{A.}},
\bauthor{\bsnm{Gutierrez}, \binits{A.M.}},
\bauthor{\bsnm{Tulli}, \binits{D.}},
\bauthor{\bsnm{Ma}, \binits{P.}},
\bauthor{\bsnm{Baeuerle}, \binits{B.}},
\bauthor{\bsnm{Josten}, \binits{A.}},
\bauthor{\bsnm{Heni}, \binits{W.}},
\bauthor{\bsnm{Caimi}, \binits{D.}},
\bauthor{\bsnm{Czornomaz}, \binits{L.}},
\bauthor{\bsnm{Demkov}, \binits{A.A.}},
\bauthor{\bsnm{Leuthold}, \binits{J.}},
\bauthor{\bsnm{Sanchis}, \binits{P.}},
\bauthor{\bsnm{Fompeyrine}, \binits{J.}}:
\batitle{Large pockels effect in micro- and nanostructured barium titanate integrated on silicon}.
\bjtitle{Nature Materials}
\bvolume{18},
\bfpage{42}--\blpage{47}
(\byear{2019})
\doiurl{10.1038/s41563-018-0208-0}
\end{barticle}
\endbibitem

\bibitem[\protect\citeauthoryear{Li et~al.}{2020}]{Li2020a}
\begin{barticle}
\bauthor{\bsnm{Li}, \binits{M.}},
\bauthor{\bsnm{Ling}, \binits{J.}},
\bauthor{\bsnm{He}, \binits{Y.}},
\bauthor{\bsnm{Javid}, \binits{U.A.}},
\bauthor{\bsnm{Xue}, \binits{S.}},
\bauthor{\bsnm{Lin}, \binits{Q.}}:
\batitle{Lithium niobate photonic-crystal electro-optic modulator}.
\bjtitle{Nature Communications}
\bvolume{11},
\bfpage{4123}
(\byear{2020})
\doiurl{10.1038/s41467-020-17950-7}
\end{barticle}
\endbibitem

\bibitem[\protect\citeauthoryear{Zhang et~al.}{2021}]{Zhang2021a}
\begin{barticle}
\bauthor{\bsnm{Zhang}, \binits{M.}},
\bauthor{\bsnm{Wang}, \binits{C.}},
\bauthor{\bsnm{Kharel}, \binits{P.}},
\bauthor{\bsnm{Zhu}, \binits{D.}},
\bauthor{\bsnm{Lončar}, \binits{M.}}:
\batitle{Integrated lithium niobate electro-optic modulators: when performance meets scalability}.
\bjtitle{Optica}
\bvolume{8},
\bfpage{652}
(\byear{2021})
\doiurl{10.1364/OPTICA.415762}
\end{barticle}
\endbibitem

\bibitem[\protect\citeauthoryear{Weigand et~al.}{2021}]{Weigand2021}
\begin{barticle}
\bauthor{\bsnm{Weigand}, \binits{H.}},
\bauthor{\bsnm{Vogler-Neuling}, \binits{V.V.}},
\bauthor{\bsnm{Escalé}, \binits{M.R.}},
\bauthor{\bsnm{Pohl}, \binits{D.}},
\bauthor{\bsnm{Richter}, \binits{F.U.}},
\bauthor{\bsnm{Karvounis}, \binits{A.}},
\bauthor{\bsnm{Timpu}, \binits{F.}},
\bauthor{\bsnm{Grange}, \binits{R.}}:
\batitle{Enhanced electro-optic modulation in resonant metasurfaces of lithium niobate}.
\bjtitle{ACS Photonics}
\bvolume{8},
\bfpage{3004}--\blpage{3009}
(\byear{2021})
\doiurl{10.1021/acsphotonics.1c00935}
\end{barticle}
\endbibitem

\bibitem[\protect\citeauthoryear{Weiss et~al.}{2022}]{Weiss2022}
\begin{barticle}
\bauthor{\bsnm{Weiss}, \binits{A.}},
\bauthor{\bsnm{Frydendahl}, \binits{C.}},
\bauthor{\bsnm{Bar-David}, \binits{J.}},
\bauthor{\bsnm{Zektzer}, \binits{R.}},
\bauthor{\bsnm{Edrei}, \binits{E.}},
\bauthor{\bsnm{Engelberg}, \binits{J.}},
\bauthor{\bsnm{Mazurski}, \binits{N.}},
\bauthor{\bsnm{Desiatov}, \binits{B.}},
\bauthor{\bsnm{Levy}, \binits{U.}}:
\batitle{Tunable metasurface using thin-film lithium niobate in the telecom regime}.
\bjtitle{ACS Photonics}
\bvolume{9},
\bfpage{605}--\blpage{612}
(\byear{2022})
\doiurl{10.1021/acsphotonics.1c01582}
\end{barticle}
\endbibitem

\bibitem[\protect\citeauthoryear{Damgaard-Carstensen and Bozhevolnyi}{2023}]{Damgaard-Carstensen2023}
\begin{barticle}
\bauthor{\bsnm{Damgaard-Carstensen}, \binits{C.}},
\bauthor{\bsnm{Bozhevolnyi}, \binits{S.I.}}:
\batitle{Nonlocal electro-optic metasurfaces for free-space light modulation}.
\bjtitle{Nanophotonics}
\bvolume{12},
\bfpage{2953}--\blpage{2962}
(\byear{2023})
\doiurl{10.1515/nanoph-2023-0042}
\end{barticle}
\endbibitem

\bibitem[\protect\citeauthoryear{Zhang et~al.}{2019}]{Zhang2019a}
\begin{barticle}
\bauthor{\bsnm{Zhang}, \binits{M.}},
\bauthor{\bsnm{Buscaino}, \binits{B.}},
\bauthor{\bsnm{Wang}, \binits{C.}},
\bauthor{\bsnm{Shams-Ansari}, \binits{A.}},
\bauthor{\bsnm{Reimer}, \binits{C.}},
\bauthor{\bsnm{Zhu}, \binits{R.}},
\bauthor{\bsnm{Kahn}, \binits{J.M.}},
\bauthor{\bsnm{Lončar}, \binits{M.}}:
\batitle{Broadband electro-optic frequency comb generation in a lithium niobate microring resonator}.
\bjtitle{Nature}
\bvolume{568},
\bfpage{373}--\blpage{377}
(\byear{2019})
\doiurl{10.1038/s41586-019-1008-7}
\end{barticle}
\endbibitem

\bibitem[\protect\citeauthoryear{Stokowski et~al.}{2024}]{Stokowski2024}
\begin{barticle}
\bauthor{\bsnm{Stokowski}, \binits{H.S.}},
\bauthor{\bsnm{Dean}, \binits{D.J.}},
\bauthor{\bsnm{Hwang}, \binits{A.Y.}},
\bauthor{\bsnm{Park}, \binits{T.}},
\bauthor{\bsnm{Celik}, \binits{O.T.}},
\bauthor{\bsnm{McKenna}, \binits{T.P.}},
\bauthor{\bsnm{Jankowski}, \binits{M.}},
\bauthor{\bsnm{Langrock}, \binits{C.}},
\bauthor{\bsnm{Ansari}, \binits{V.}},
\bauthor{\bsnm{Fejer}, \binits{M.M.}},
\bauthor{\bsnm{Safavi-Naeini}, \binits{A.H.}}:
\batitle{Integrated frequency-modulated optical parametric oscillator}.
\bjtitle{Nature}
\bvolume{627},
\bfpage{95}--\blpage{100}
(\byear{2024})
\doiurl{10.1038/s41586-024-07071-2}
\end{barticle}
\endbibitem

\bibitem[\protect\citeauthoryear{Fedotova et~al.}{2022}]{Fedotova2022}
\begin{barticle}
\bauthor{\bsnm{Fedotova}, \binits{A.}},
\bauthor{\bsnm{Carletti}, \binits{L.}},
\bauthor{\bsnm{Zilli}, \binits{A.}},
\bauthor{\bsnm{Setzpfandt}, \binits{F.}},
\bauthor{\bsnm{Staude}, \binits{I.}},
\bauthor{\bsnm{Toma}, \binits{A.}},
\bauthor{\bsnm{Finazzi}, \binits{M.}},
\bauthor{\bsnm{Angelis}, \binits{C.D.}},
\bauthor{\bsnm{Pertsch}, \binits{T.}},
\bauthor{\bsnm{Neshev}, \binits{D.N.}},
\bauthor{\bsnm{Celebrano}, \binits{M.}}:
\batitle{Lithium niobate meta-optics}.
\bjtitle{ACS Photonics}
\bvolume{9},
\bfpage{3745}--\blpage{3763}
(\byear{2022})
\doiurl{10.1021/acsphotonics.2c00835}
\end{barticle}
\endbibitem

\bibitem[\protect\citeauthoryear{Chen et~al.}{2025}]{Chen2025}
\begin{barticle}
\bauthor{\bsnm{Chen}, \binits{Z.}},
\bauthor{\bsnm{Mazurski}, \binits{N.}},
\bauthor{\bsnm{Engelberg}, \binits{J.}},
\bauthor{\bsnm{Levy}, \binits{U.}}:
\batitle{Tunable transmissive metasurface based on thin-film lithium niobate}.
\bjtitle{ACS Photonics}
\bvolume{12},
\bfpage{1174}--\blpage{1183}
(\byear{2025})
\doiurl{10.1021/acsphotonics.4c02354}
\end{barticle}
\endbibitem

\bibitem[\protect\citeauthoryear{Kharel et~al.}{2021}]{Kharel2021}
\begin{barticle}
\bauthor{\bsnm{Kharel}, \binits{P.}},
\bauthor{\bsnm{Reimer}, \binits{C.}},
\bauthor{\bsnm{Luke}, \binits{K.}},
\bauthor{\bsnm{He}, \binits{L.}},
\bauthor{\bsnm{Zhang}, \binits{M.}}:
\batitle{Breaking voltage–bandwidth limits in integrated lithium niobate modulators using micro-structured electrodes}.
\bjtitle{Optica}
\bvolume{8},
\bfpage{357}
(\byear{2021})
\doiurl{10.1364/OPTICA.416155}
\end{barticle}
\endbibitem

\bibitem[\protect\citeauthoryear{Li et~al.}{2023}]{Li2023b}
\begin{barticle}
\bauthor{\bsnm{Li}, \binits{Y.}},
\bauthor{\bsnm{Lan}, \binits{T.}},
\bauthor{\bsnm{Yang}, \binits{D.}},
\bauthor{\bsnm{Bao}, \binits{J.}},
\bauthor{\bsnm{Xiang}, \binits{M.}},
\bauthor{\bsnm{Yang}, \binits{F.}},
\bauthor{\bsnm{Wang}, \binits{Z.}}:
\batitle{High-performance mach–zehnder modulator based on thin-film lithium niobate with low voltage-length product}.
\bjtitle{ACS Omega}
\bvolume{8},
\bfpage{9644}--\blpage{9651}
(\byear{2023})
\doiurl{10.1021/acsomega.3c00310}
\end{barticle}
\endbibitem

\bibitem[\protect\citeauthoryear{Mercante et~al.}{2016}]{Mercante2016}
\begin{barticle}
\bauthor{\bsnm{Mercante}, \binits{A.J.}},
\bauthor{\bsnm{Yao}, \binits{P.}},
\bauthor{\bsnm{Shi}, \binits{S.}},
\bauthor{\bsnm{Schneider}, \binits{G.}},
\bauthor{\bsnm{Murakowski}, \binits{J.}},
\bauthor{\bsnm{Prather}, \binits{D.W.}}:
\batitle{110 ghz cmos compatible thin film linbo3 modulator on silicon}.
\bjtitle{Optics Express}
\bvolume{24},
\bfpage{15590}
(\byear{2016})
\doiurl{10.1364/OE.24.015590}
\end{barticle}
\endbibitem

\bibitem[\protect\citeauthoryear{Barton et~al.}{2020}]{Barton2020}
\begin{barticle}
\bauthor{\bsnm{Barton}, \binits{D.}},
\bauthor{\bsnm{Hu}, \binits{J.}},
\bauthor{\bsnm{Dixon}, \binits{J.}},
\bauthor{\bsnm{Klopfer}, \binits{E.}},
\bauthor{\bsnm{Dagli}, \binits{S.}},
\bauthor{\bsnm{Lawrence}, \binits{M.}},
\bauthor{\bsnm{Dionne}, \binits{J.}}:
\batitle{High-q nanophotonics: sculpting wavefronts with slow light}.
\bjtitle{Nanophotonics}
\bvolume{10},
\bfpage{83}--\blpage{88}
(\byear{2020})
\doiurl{10.1515/nanoph-2020-0510}
\end{barticle}
\endbibitem

\bibitem[\protect\citeauthoryear{Hail et~al.}{2023}]{Hail2023}
\begin{barticle}
\bauthor{\bsnm{Hail}, \binits{C.U.}},
\bauthor{\bsnm{Foley}, \binits{M.}},
\bauthor{\bsnm{Sokhoyan}, \binits{R.}},
\bauthor{\bsnm{Michaeli}, \binits{L.}},
\bauthor{\bsnm{Atwater}, \binits{H.A.}}:
\batitle{High quality factor metasurfaces for two-dimensional wavefront manipulation}.
\bjtitle{Nature Communications}
\bvolume{14},
\bfpage{8476}
(\byear{2023})
\doiurl{10.1038/s41467-023-44164-4}
\end{barticle}
\endbibitem

\bibitem[\protect\citeauthoryear{So et~al.}{2023}]{So2023}
\begin{botherref}
\oauthor{\bsnm{So}, \binits{S.}},
\oauthor{\bsnm{Mun}, \binits{J.}},
\oauthor{\bsnm{Park}, \binits{J.}},
\oauthor{\bsnm{Rho}, \binits{J.}}:
Revisiting the design strategies for metasurfaces: Fundamental physics, optimization, and beyond.
Advanced Materials
\textbf{35}
(2023)
\doiurl{10.1002/adma.202206399}
\end{botherref}
\endbibitem

\bibitem[\protect\citeauthoryear{Francescantonio et~al.}{2024}]{DiFrancescantonio2024}
\begin{botherref}
\oauthor{\bsnm{Francescantonio}, \binits{A.D.}},
\oauthor{\bsnm{Sabatti}, \binits{A.}},
\oauthor{\bsnm{Weigand}, \binits{H.}},
\oauthor{\bsnm{Bailly}, \binits{E.}},
\oauthor{\bsnm{Vincenti}, \binits{M.A.}},
\oauthor{\bsnm{Carletti}, \binits{L.}},
\oauthor{\bsnm{Kellner}, \binits{J.}},
\oauthor{\bsnm{Zilli}, \binits{A.}},
\oauthor{\bsnm{Finazzi}, \binits{M.}},
\oauthor{\bsnm{Celebrano}, \binits{M.}},
\oauthor{\bsnm{Grange}, \binits{R.}}:
Efficient ghz electro-optical modulation with a nonlocal lithium niobate metasurface in the linear and nonlinear regime.
arXiv preprint arXiv:2412.03422
(2024)
\end{botherref}
\endbibitem

\bibitem[\protect\citeauthoryear{Damgaard-Carstensen et~al.}{2024}]{Damgaard-Carstensen2024}
\begin{botherref}
\oauthor{\bsnm{Damgaard-Carstensen}, \binits{C.}},
\oauthor{\bsnm{Yezekyan}, \binits{T.}},
\oauthor{\bsnm{Brongersma}, \binits{M.L.}},
\oauthor{\bsnm{Bozhevolnyi}, \binits{S.I.}}:
Highly efficient, tunable, electro-optic metasurfaces based on quasi-bound states in the continuum.
arXiv preprint arXiv:2412.01449
(2024)
\end{botherref}
\endbibitem

\bibitem[\protect\citeauthoryear{Overvig et~al.}{2020}]{Overvig2020}
\begin{barticle}
\bauthor{\bsnm{Overvig}, \binits{A.C.}},
\bauthor{\bsnm{Malek}, \binits{S.C.}},
\bauthor{\bsnm{Carter}, \binits{M.J.}},
\bauthor{\bsnm{Shrestha}, \binits{S.}},
\bauthor{\bsnm{Yu}, \binits{N.}}:
\batitle{Selection rules for quasibound states in the continuum}.
\bjtitle{Physical Review B}
\bvolume{102},
\bfpage{035434}
(\byear{2020})
\doiurl{10.1103/PhysRevB.102.035434}
\end{barticle}
\endbibitem

\bibitem[\protect\citeauthoryear{Shastri and Monticone}{2023}]{Shastri2023}
\begin{barticle}
\bauthor{\bsnm{Shastri}, \binits{K.}},
\bauthor{\bsnm{Monticone}, \binits{F.}}:
\batitle{Nonlocal flat optics}.
\bjtitle{Nature Photonics}
\bvolume{17},
\bfpage{36}--\blpage{47}
(\byear{2023})
\doiurl{10.1038/s41566-022-01098-5}
\end{barticle}
\endbibitem

\bibitem[\protect\citeauthoryear{Koshelev et~al.}{2018}]{Koshelev2018b}
\begin{barticle}
\bauthor{\bsnm{Koshelev}, \binits{K.}},
\bauthor{\bsnm{Lepeshov}, \binits{S.}},
\bauthor{\bsnm{Liu}, \binits{M.}},
\bauthor{\bsnm{Bogdanov}, \binits{A.}},
\bauthor{\bsnm{Kivshar}, \binits{Y.}}:
\batitle{Asymmetric metasurfaces with high-q resonances governed by bound states in the continuum}.
\bjtitle{Physical Review Letters}
\bvolume{121},
\bfpage{193903}
(\byear{2018})
\doiurl{10.1103/PhysRevLett.121.193903}
\end{barticle}
\endbibitem

\bibitem[\protect\citeauthoryear{Witmer et~al.}{2017}]{Witmer2017d}
\begin{barticle}
\bauthor{\bsnm{Witmer}, \binits{J.D.}},
\bauthor{\bsnm{Valery}, \binits{J.A.}},
\bauthor{\bsnm{Arrangoiz-Arriola}, \binits{P.}},
\bauthor{\bsnm{Sarabalis}, \binits{C.J.}},
\bauthor{\bsnm{Hill}, \binits{J.T.}},
\bauthor{\bsnm{Safavi-Naeini}, \binits{A.H.}}:
\batitle{High-q photonic resonators and electro-optic coupling using silicon-on-lithium-niobate}.
\bjtitle{Scientific Reports}
\bvolume{7},
\bfpage{46313}
(\byear{2017})
\doiurl{10.1038/srep46313}
\end{barticle}
\endbibitem

\bibitem[\protect\citeauthoryear{Weigel et~al.}{2016}]{Weigel2016d}
\begin{barticle}
\bauthor{\bsnm{Weigel}, \binits{P.O.}},
\bauthor{\bsnm{Savanier}, \binits{M.}},
\bauthor{\bsnm{DeRose}, \binits{C.T.}},
\bauthor{\bsnm{Pomerene}, \binits{A.T.}},
\bauthor{\bsnm{Starbuck}, \binits{A.L.}},
\bauthor{\bsnm{Lentine}, \binits{A.L.}},
\bauthor{\bsnm{Stenger}, \binits{V.}},
\bauthor{\bsnm{Mookherjea}, \binits{S.}}:
\batitle{Lightwave circuits in lithium niobate through hybrid waveguides with silicon photonics}.
\bjtitle{Scientific Reports}
\bvolume{6},
\bfpage{22301}
(\byear{2016})
\doiurl{10.1038/srep22301}
\end{barticle}
\endbibitem

\bibitem[\protect\citeauthoryear{Chiles and Fathpour}{2014}]{Chiles2014b}
\begin{barticle}
\bauthor{\bsnm{Chiles}, \binits{J.}},
\bauthor{\bsnm{Fathpour}, \binits{S.}}:
\batitle{Mid-infrared integrated waveguide modulators based on silicon-on-lithium-niobate photonics}.
\bjtitle{Optica}
\bvolume{1},
\bfpage{350}
(\byear{2014})
\doiurl{10.1364/OPTICA.1.000350}
\end{barticle}
\endbibitem

\bibitem[\protect\citeauthoryear{Klopfer et~al.}{2022}]{Klopfer2022}
\begin{barticle}
\bauthor{\bsnm{Klopfer}, \binits{E.}},
\bauthor{\bsnm{Dagli}, \binits{S.}},
\bauthor{\bsnm{Barton}, \binits{D.}},
\bauthor{\bsnm{Lawrence}, \binits{M.}},
\bauthor{\bsnm{Dionne}, \binits{J.A.}}:
\batitle{High-quality-factor silicon-on-lithium niobate metasurfaces for electro-optically reconfigurable wavefront shaping}.
\bjtitle{Nano Letters}
\bvolume{22},
\bfpage{1703}--\blpage{1709}
(\byear{2022})
\doiurl{10.1021/acs.nanolett.1c04723}
\end{barticle}
\endbibitem

\bibitem[\protect\citeauthoryear{Chelladurai et~al.}{2024}]{Chelladurai2024}
\begin{botherref}
\oauthor{\bsnm{Chelladurai}, \binits{D.}},
\oauthor{\bsnm{Kohli}, \binits{M.}},
\oauthor{\bsnm{Winiger}, \binits{J.}},
\oauthor{\bsnm{Moor}, \binits{D.}},
\oauthor{\bsnm{Messner}, \binits{A.}},
\oauthor{\bsnm{Fedoryshyn}, \binits{Y.}},
\oauthor{\bsnm{Eleraky}, \binits{M.}},
\oauthor{\bsnm{Liu}, \binits{Y.}},
\oauthor{\bsnm{Wang}, \binits{H.}},
\oauthor{\bsnm{Leuthold}, \binits{J.}}:
Barium titanate and lithium niobate permittivity and pockels coefficients from mhz to sub-thz frequencies.
arXiv preprint arXiv:2407.03443
(2024)
\end{botherref}
\endbibitem

\bibitem[\protect\citeauthoryear{Witmer et~al.}{2016}]{Witmer2016d}
\begin{barticle}
\bauthor{\bsnm{Witmer}, \binits{J.D.}},
\bauthor{\bsnm{Hill}, \binits{J.T.}},
\bauthor{\bsnm{Safavi-Naeini}, \binits{A.H.}}:
\batitle{Design of nanobeam photonic crystal resonators for a silicon-on-lithium-niobate platform}.
\bjtitle{Optics Express}
\bvolume{24},
\bfpage{5876}
(\byear{2016})
\doiurl{10.1364/OE.24.005876}
\end{barticle}
\endbibitem

\bibitem[\protect\citeauthoryear{Barton et~al.}{2021}]{Barton2021d}
\begin{barticle}
\bauthor{\bsnm{Barton}, \binits{D.}},
\bauthor{\bsnm{Lawrence}, \binits{M.}},
\bauthor{\bsnm{Dionne}, \binits{J.}}:
\batitle{Wavefront shaping and modulation with resonant electro-optic phase gradient metasurfaces}.
\bjtitle{Applied Physics Letters}
\bvolume{118},
\bfpage{071104}
(\byear{2021})
\doiurl{10.1063/5.0039873}
\end{barticle}
\endbibitem

\bibitem[\protect\citeauthoryear{Lawrence et~al.}{2018}]{Lawrence2018i}
\begin{barticle}
\bauthor{\bsnm{Lawrence}, \binits{M.}},
\bauthor{\bsnm{Barton}, \binits{D.R.}},
\bauthor{\bsnm{Dionne}, \binits{J.A.}}:
\batitle{{Nonreciprocal Flat Optics with Silicon Metasurfaces}}.
\bjtitle{Nano Letters}
\bvolume{18}(\bissue{2}),
\bfpage{1104}--\blpage{1109}
(\byear{2018})
\doiurl{10.1021/acs.nanolett.7b04646}
\end{barticle}
\endbibitem

\bibitem[\protect\citeauthoryear{Overvig et~al.}{2018}]{Overvig2018}
\begin{barticle}
\bauthor{\bsnm{Overvig}, \binits{A.C.}},
\bauthor{\bsnm{Shrestha}, \binits{S.}},
\bauthor{\bsnm{Yu}, \binits{N.}}:
\batitle{Dimerized high contrast gratings}.
\bjtitle{Nanophotonics}
\bvolume{7},
\bfpage{1157}--\blpage{1168}
(\byear{2018})
\doiurl{10.1515/nanoph-2017-0127}
\end{barticle}
\endbibitem

\bibitem[\protect\citeauthoryear{Kim et~al.}{2019}]{Kim2019k}
\begin{barticle}
\bauthor{\bsnm{Kim}, \binits{S.}},
\bauthor{\bsnm{Kim}, \binits{K.-H.}},
\bauthor{\bsnm{Cahoon}, \binits{J.}}:
\batitle{Optical bound states in the continuum with nanowire geometric superlattices}.
\bjtitle{Physical Review Letters}
\bvolume{122},
\bfpage{187402}
(\byear{2019})
\doiurl{10.1103/PhysRevLett.122.187402}
\end{barticle}
\endbibitem

\bibitem[\protect\citeauthoryear{Dolia et~al.}{2024}]{Dolia2024}
\begin{barticle}
\bauthor{\bsnm{Dolia}, \binits{V.}},
\bauthor{\bsnm{Balch}, \binits{H.B.}},
\bauthor{\bsnm{Dagli}, \binits{S.}},
\bauthor{\bsnm{Abdollahramezani}, \binits{S.}},
\bauthor{\bsnm{Delgado}, \binits{H.C.}},
\bauthor{\bsnm{Moradifar}, \binits{P.}},
\bauthor{\bsnm{Chang}, \binits{K.}},
\bauthor{\bsnm{Stiber}, \binits{A.}},
\bauthor{\bsnm{Safir}, \binits{F.}},
\bauthor{\bsnm{Lawrence}, \binits{M.}},
\bauthor{\bsnm{Hu}, \binits{J.}},
\bauthor{\bsnm{Dionne}, \binits{J.A.}}:
\batitle{Very-large-scale-integrated high quality factor nanoantenna pixels}.
\bjtitle{Nature Nanotechnology}
\bvolume{19},
\bfpage{1290}--\blpage{1298}
(\byear{2024})
\doiurl{10.1038/s41565-024-01697-z}
\end{barticle}
\endbibitem

\bibitem[\protect\citeauthoryear{Lawrence et~al.}{2020}]{Lawrence2020d}
\begin{barticle}
\bauthor{\bsnm{Lawrence}, \binits{M.}},
\bauthor{\bsnm{Barton}, \binits{D.R.}},
\bauthor{\bsnm{Dixon}, \binits{J.}},
\bauthor{\bsnm{Song}, \binits{J.-H.}},
\bauthor{\bsnm{Groep}, \binits{J.}},
\bauthor{\bsnm{Brongersma}, \binits{M.L.}},
\bauthor{\bsnm{Dionne}, \binits{J.A.}}:
\batitle{High quality factor phase gradient metasurfaces}.
\bjtitle{Nature Nanotechnology}
\bvolume{15},
\bfpage{956}--\blpage{961}
(\byear{2020})
\doiurl{10.1038/s41565-020-0754-x}
\end{barticle}
\endbibitem

\bibitem[\protect\citeauthoryear{Lin et~al.}{2023}]{Lin2023}
\begin{barticle}
\bauthor{\bsnm{Lin}, \binits{L.}},
\bauthor{\bsnm{Hu}, \binits{J.}},
\bauthor{\bsnm{Dagli}, \binits{S.}},
\bauthor{\bsnm{Dionne}, \binits{J.A.}},
\bauthor{\bsnm{Lawrence}, \binits{M.}}:
\batitle{Universal narrowband wavefront shaping with high quality factor meta-reflect-arrays}.
\bjtitle{Nano Letters}
\bvolume{23},
\bfpage{1355}--\blpage{1362}
(\byear{2023})
\doiurl{10.1021/acs.nanolett.2c04621}
\end{barticle}
\endbibitem

\bibitem[\protect\citeauthoryear{Klopfer et~al.}{2023}]{Klopfer2023}
\begin{botherref}
\oauthor{\bsnm{Klopfer}, \binits{E.}},
\oauthor{\bsnm{Delgado}, \binits{H.C.}},
\oauthor{\bsnm{Dagli}, \binits{S.}},
\oauthor{\bsnm{Lawrence}, \binits{M.}},
\oauthor{\bsnm{Dionne}, \binits{J.A.}}:
A thermally controlled high-q metasurface lens.
Applied Physics Letters
\textbf{122}
(2023)
\doiurl{10.1063/5.0152535}
\end{botherref}
\endbibitem

\bibitem[\protect\citeauthoryear{Limonov et~al.}{2017}]{Limonov2017b}
\begin{barticle}
\bauthor{\bsnm{Limonov}, \binits{M.F.}},
\bauthor{\bsnm{Rybin}, \binits{M.V.}},
\bauthor{\bsnm{Poddubny}, \binits{A.N.}},
\bauthor{\bsnm{Kivshar}, \binits{Y.S.}}:
\batitle{{Fano resonances in photonics}}.
\bjtitle{Nature Photonics}
\bvolume{11}(\bissue{9}),
\bfpage{543}--\blpage{554}
(\byear{2017})
\doiurl{10.1038/NPHOTON.2017.142}
\end{barticle}
\endbibitem

\bibitem[\protect\citeauthoryear{Sisler et~al.}{2024}]{Sisler2024}
\begin{barticle}
\bauthor{\bsnm{Sisler}, \binits{J.}},
\bauthor{\bsnm{Thureja}, \binits{P.}},
\bauthor{\bsnm{Grajower}, \binits{M.Y.}},
\bauthor{\bsnm{Sokhoyan}, \binits{R.}},
\bauthor{\bsnm{Huang}, \binits{I.}},
\bauthor{\bsnm{Atwater}, \binits{H.A.}}:
\batitle{Electrically tunable space–time metasurfaces at optical frequencies}.
\bjtitle{Nature Nanotechnology}
\bvolume{19},
\bfpage{1491}--\blpage{1498}
(\byear{2024})
\doiurl{10.1038/s41565-024-01728-9}
\end{barticle}
\endbibitem

\bibitem[\protect\citeauthoryear{Jiang et~al.}{2020}]{Jiang2020}
\begin{barticle}
\bauthor{\bsnm{Jiang}, \binits{W.}},
\bauthor{\bsnm{Sarabalis}, \binits{C.J.}},
\bauthor{\bsnm{Dahmani}, \binits{Y.D.}},
\bauthor{\bsnm{Patel}, \binits{R.N.}},
\bauthor{\bsnm{Mayor}, \binits{F.M.}},
\bauthor{\bsnm{McKenna}, \binits{T.P.}},
\bauthor{\bsnm{Laer}, \binits{R.V.}},
\bauthor{\bsnm{Safavi-Naeini}, \binits{A.H.}}:
\batitle{Efficient bidirectional piezo-optomechanical transduction between microwave and optical frequency}.
\bjtitle{Nature Communications}
\bvolume{11},
\bfpage{1166}
(\byear{2020})
\doiurl{10.1038/s41467-020-14863-3}
\end{barticle}
\endbibitem

\bibitem[\protect\citeauthoryear{Santiago-Cruz et~al.}{2021}]{Santiago-Cruz2021}
\begin{barticle}
\bauthor{\bsnm{Santiago-Cruz}, \binits{T.}},
\bauthor{\bsnm{Fedotova}, \binits{A.}},
\bauthor{\bsnm{Sultanov}, \binits{V.}},
\bauthor{\bsnm{Weissflog}, \binits{M.A.}},
\bauthor{\bsnm{Arslan}, \binits{D.}},
\bauthor{\bsnm{Younesi}, \binits{M.}},
\bauthor{\bsnm{Pertsch}, \binits{T.}},
\bauthor{\bsnm{Staude}, \binits{I.}},
\bauthor{\bsnm{Setzpfandt}, \binits{F.}},
\bauthor{\bsnm{Chekhova}, \binits{M.}}:
\batitle{Photon pairs from resonant metasurfaces}.
\bjtitle{Nano Letters}
\bvolume{21},
\bfpage{4423}--\blpage{4429}
(\byear{2021})
\doiurl{10.1021/acs.nanolett.1c01125}
\end{barticle}
\endbibitem

\bibitem[\protect\citeauthoryear{Zhang et~al.}{2022}]{Zhang2022}
\begin{barticle}
\bauthor{\bsnm{Zhang}, \binits{J.}},
\bauthor{\bsnm{Ma}, \binits{J.}},
\bauthor{\bsnm{Parry}, \binits{M.}},
\bauthor{\bsnm{Cai}, \binits{M.}},
\bauthor{\bsnm{Camacho-Morales}, \binits{R.}},
\bauthor{\bsnm{Xu}, \binits{L.}},
\bauthor{\bsnm{Neshev}, \binits{D.N.}},
\bauthor{\bsnm{Sukhorukov}, \binits{A.A.}}:
\batitle{Spatially entangled photon pairs from lithium niobate nonlocal metasurfaces}.
\bjtitle{Science Advances}
\bvolume{8},
\bfpage{4240}
(\byear{2022})
\doiurl{10.1126/sciadv.abq4240}
\end{barticle}
\endbibitem

\end{thebibliography}

\end{document}